\documentclass[iop]{emulateapj}

\usepackage{amsmath}
\usepackage{color}

\shorttitle{AGN spectra for photoionization modeling}
\shortauthors{A. D. Thomas et al.}

\begin{document}

\title{A physically-based model of the ionizing radiation from active galaxies for photoionization modeling}

\author{A. D. Thomas}
\author{B. A. Groves}
\author{R. S. Sutherland}
\author{M. A. Dopita}
\author{L. J. Kewley}
\affil{Research School of Astronomy and Astrophysics, Australian National University, Canberra, ACT 2611, Australia} 
\email{adam.thomas@anu.edu.au}

\and

\author{C. Jin}
\affil{Max-Planck-Institut f\"{u}r Extraterrestrische Physik, Giessenbachstrasse, D-85748 Garching, Germany}
\email{chichuan@mpe.mpg.de}

\begin{abstract}
We present a simplified model of Active Galactic Nucleus (AGN) continuum emission designed for photoionization modeling.  The new model {\sc oxaf} reproduces the diversity of spectral shapes that arise in physically-based models.  We identify and explain degeneracies in the effects of AGN parameters on model spectral shapes, with a focus on the complete degeneracy between the black hole mass and AGN luminosity.  Our re-parametrized model {\sc oxaf} removes these degeneracies and accepts three parameters which directly describe the output spectral shape: the energy of the peak of the accretion disk emission $E_{\mathrm{peak}}$, the photon power-law index of the non-thermal emission $\Gamma$, and the proportion of the total flux which is emitted in the non-thermal component $p_{\mathrm{NT}}$. The parameter $E_{\mathrm{peak}}$ is presented as a function of the black hole mass, AGN luminosity, and `coronal radius' of the {\sc optxagnf} model upon which {\sc oxaf} is based.  We show that the soft X-ray excess does not significantly affect photoionization modeling predictions of strong emission lines in Seyfert narrow-line regions.  Despite its simplicity, {\sc oxaf} accounts for opacity effects where the accretion disk is ionized because it inherits the `color correction' of {\sc optxagnf}. We use a grid of {\sc mappings} photoionization models with {\sc oxaf} ionizing spectra to demonstrate how predicted emission-line ratios on standard optical diagnostic diagrams are sensitive to each of the three {\sc oxaf} parameters.  The {\sc oxaf} code is publicly available in the Astrophysics Source Code Library.
\end{abstract}


\section{Introduction}

Active galactic nuclei (AGN) feature narrow-line regions (NLRs) in which continuum radiation from the central engine photoionizes the surrounding interstellar medium (ISM).  The relatively hard AGN continuum emission injects more energy and photoionizes the ISM to higher ionization states than the ionizing spectra of OB stars at the centers of \ion{H}{2} regions. 

Many insights into the physics of the the ISM in active galaxies have been gained through the use of photoionization models.  The results of photoionization modeling facilitate the routine diagnosis of the density, temperature, and cause of ionization of NLR gas from observed emission-line spectra.  Photoionization models have been used to explain the very similar NLR ionization parameters observed across the population of active galaxies - this effect is caused by the gas pressure gradient balancing the radiation pressure gradient in dusty NLR clouds such that the `local' ionization parameter is self-regulated \citep{Dopita_2002_Dusty_NLRs, Groves_2004_NLR_I}.  Models are an important tool in investigating the effects of various ISM parameters on emission-line spectra; e.g.\ determining the effects of changing abundances \citep{Groves_2006_low_Z_AGN}.  Recently, \citet{Davies_2016_S7_NLR_rad_P} showed how comparisons between models and spatially-resolved spectroscopy may reveal where extended NLRs are radiation pressure-dominated or gas pressure-dominated on scales of tens to hundreds of parsecs.

The emission line spectra of NLRs in active galaxies strongly depend upon the shape of the spectrum of the central source at energies immediately above the Lyman limit at 13.6~eV.  These ionizing photons are not directly observable because they are very strongly absorbed in the NLR and broader ISM of the host galaxy, as well as in the ISM of the Milky Way.

Up to the present, ionizing AGN spectra for photoionization modeling have almost exclusively been constructed using piecewise functions of one to three simple power-laws \citep[e.g.][]{Viegas-Aldrovandi_Contini_1989, Murayama_Taniguchi_1998, Allen_1998_NLRs_UV, Evans_1999, Contini_Viegas_2001, Collins_2009}.  \citet{Groves_2004_NLR_I} use a conventional range of power-law indices of $\alpha = -2.0$ to $\alpha = -1.2$, for a single power-law covering 0.005 to 1~keV.  The authors find that a steeper power-law leads to weaker high-ionization lines and relatively stronger H lines, due to the relative excess of H- but not He-ionizing photons.  A more physically-motivated empirical two-component model for the ionizing spectrum is used, for example, in \citet{Bland-Hawthorn_2013_MW_centre}).

In this work we seek to develop a modern AGN continuum emission model for use as the input spectrum in photoionization models, based on physically-derived models, and including for example ionization effects in the shape of the accretion disk spectrum.  We aim to identify the important AGN parameters which influence the shape of the ionizing spectrum and hence the predicted emission-line ratios.  This allows us to minimize the parameter space used to cover a large range of model spectra.  A small parameter space is easier to handle conceptually and computationally, and the process of combining fully and partially degenerate parameters into a smaller set of parameters results in insights into how the shape of the ionizing spectrum may arise from various combinations of fundamental AGN parameters.

In photoionization modeling we assume that the model ionizing spectrum will be intrinsic when it reaches the modeled clouds, i.e.\ that there is no significant absorption between the central engine and the modeled clouds.  Hence the spectra will primarily feature a peak formed by a superposition of pseudo-blackbodies due to the accretion disk, and a power-law non-thermal component resulting from Compton up-scattering of disk photons.  A physical model of these spectra may be expected to be well-described by few parameters.

A detailed model of continuum emission from AGN must rely on theoretical emissivities from accretion disk models.  The standard Novikov-Thorne `thin disk' accretion model (\citet{Novikov_Thorne_1973}, building on the work of \citet{Shakura_Sunyaev_1973}) is applicable over much of the accretion rate regime of Seyfert galaxies (${\sim}0.05$ to ${\sim}0.3$~$L/L_{\mathrm{Edd}}$).  This model has parameters such as the SMBH mass, spin, and accretion rate, which influence the theoretical ionizing AGN spectrum.

Understanding how the key parameters control the shape of the ionizing EUV-soft X-ray spectrum of AGN allows us to produce a model which keeps the number of free parameters to a minimum. Our simplified spectral modeling is, however, based firmly on the {\sc optxagnf} model \citep{Done_2012_AGN_SED, Jin_2012_I, Jin_2012_II, Jin_2012_III}.  We chose this model because it is a recent, widely-used model, it calculates theoretical spectra using physics such as the Novikov-Thorne thin accretion disk model, and because it uses a `color temperature correction' which accounts for electron-scattering opacity in parts of the disk where hydrogen is ionized.  The {\sc optxagnf} model provides the basis for a new physically-based model of the ionizing AGN spectrum for photoionization modeling.

In photoionization modeling, the ionization parameter and density are important input parameters, so the flux incident on the model cloud is easily calculated.  However the actual luminosity of the ionizing source is important only if the modeler is interested in determining the distance of the modeled cloud from the source.  We are primarily concerned with the shape of the ionizing spectrum, because this determines the relative emission-line fluxes.  Any parameter redundancies may render more difficult the interpretation of the NLR spectrum caused by a given AGN ionizing spectrum.  In this work we identify and explain various spectral-shape degeneracies, such as the complete degeneracy between the black hole mass and luminosity in the Novikov-Thorne emissivity, and successfully remove the degeneracies by constructing a new model.

We simplify the {\sc optxagnf} model by carefully removing and merging degenerate parameters.  The resulting model, {\sc oxaf}, satisfies our requirements by generating realistic model spectral shapes using the fewest possible parameters, and with all of the parameters having a direct impact on the shape of the model spectrum.  We then investigate the effect of the {\sc oxaf} parameters on photoionization models of NLR clouds.

In Section~\ref{sec:accretion_models} we describe the spectral-shape degeneracies that arise in the thin-disc accretion model and in the the {\sc optxagnf} model, and in Section~\ref{sec:param_reduction} we describe the parameters used to re-parametrize {\sc optxagnf}.  In Section~\ref{sec:oxaf} we explain how these parameters are used to construct our simplified model {\sc oxaf} and describe the implementation and validation of the new model.  In Section~\ref{sec:oxaf_predictions} we explore how the {\sc oxaf} parameters affect photoionization model predictions for diagnostic emission lines.  Our conclusions are presented in Section~\ref{sec:conclusion}.

\section{Relationships between parameters in AGN spectral models}
\label{sec:accretion_models}

The unified model of AGN describes a central, accreting supermassive black hole (SMBH) surrounded by a broad line region, a dusty obscuring torus, and a narrow-line region \citep{Antonucci_1993_AGN_unified}.  The powerful continuum radiation from the central accretion structure illuminates the broad and narrow-line regions.

Important parameters which determine the spectrum of the central accretion structure are the Eddington ratio $L/L_{\mathrm{Edd}}$, the SMBH mass $M_\mathrm{BH}$, the dimensionless SMBH spin $a^*$, and the innermost radius at which the disk is directly visible.  In this section we describe the effects of these parameters on the shape of the ionizing spectrum and the relationships between the parameters.

\subsection{The relationship between the SMBH mass and Eddington rate}
\label{sec:M_L_degeneracy}

The Novikov-Thorne thin disk model produces a clean relationship between $M_\mathrm{BH}$ and $L/L_{\textrm{Edd}}$, with these parameters having the same effect on the model spectral shape.

The temperature $T$ at a given radius of the accretion disk $r_1$ may be calculated using the following equation for the `outer' region of the disk \citep[Equation~5.10.1 in][]{Novikov_Thorne_1973}:
\begin{equation}
T(M_\mathrm{BH}, \dot{M}_\mathrm{BH}, a^*, r_1) = \left[ \frac{3 G M_\mathrm{BH} \dot{M}_\mathrm{BH}}{8 \pi \sigma_\mathrm{SB} r_1^3}  \mathcal{B}^{-1} \mathcal{C}^{-1/2} \mathcal{Q} \right]^{\frac{1}{4}}   \label{eq:T1}
\end{equation} 
where $M_\mathrm{BH}$ is the black hole mass, $\dot{M}_\mathrm{BH}$ is the accretion rate, $a^*$ is the dimensionless spin parameter, $\sigma_\mathrm{SB}$ is the Stefan-Boltzmann constant, and $G$ is the gravitational constant.  With $r$ defined as the radius normalized by the gravitational radius $r_g$ ($r_g = GM_\mathrm{BH}/c^2$; $r$ = $r_1/r_g$), the quantities $\mathcal{B}(r, a^*)$, $\mathcal{C}(r, a^*)$ and $\mathcal{Q}(r, a^*)$ are complicated dimensionless radial functions associated with relativistic corrections in the analytic accretion disk solution of \citet{Novikov_Thorne_1973} (the $\mathcal{Q}$ function was presented in the follow-up work of \citet{Page_Thorne_1974}; $\mathcal{B}$, $\mathcal{C}$ and $\mathcal{Q}$ tend to 1 at large $r$).

Using $L = \epsilon (a^*) \, \dot{M}_\mathrm{BH} \, c^2$ where $\epsilon(a^*)$ is the total accretion efficiency, and with the definition of $r$ given above, Equation~\ref{eq:T1} may be rewritten as
\begin{equation}
T(M_\mathrm{BH}, L, a^*, r) = \left[ \frac{C}{\epsilon(a^*) r^3} \frac{L}{M_\mathrm{BH}^2} \mathcal{S}(r, a^*) \right]^{\frac{1}{4}}   \label{eq:T2}
\end{equation} 
where $C$ is a collection of physical and numerical constants and $\mathcal{S}(r, a^*) = \mathcal{B}(r, a^*)^{-1} \mathcal{C}(r, a^*)^{-1/2} \mathcal{Q}(r, a^*)$.

Inverting Equation~\ref{eq:T2}, we find that
\begin{equation}
\frac{L/L_{\mathrm{Edd}}}{M_\mathrm{BH}} = F((r,T), a^*) \label{eq:ML_degen}
\end{equation} 
for a function $F$, since $L_{\mathrm{Edd}} \propto M_\mathrm{BH}$.  Specifying the temperature at a particular normalized radius as well as the spin uniquely determines the ratio $(L/L_{\mathrm{Edd}})/M_\mathrm{BH}$ such that the parameters $L/L_{\mathrm{Edd}}$ and $M_\mathrm{BH}$ are degenerate in their effects on the spectral shape.

\subsection{The {\sc optxagnf} model}
\label{sec:optxagnf}

The {\sc optxagnf} model \citep{Done_2012_AGN_SED, Jin_2012_I, Jin_2012_II, Jin_2012_III} is a significant effort in the development of theoretical models of continuum radiation from accretion onto black holes that match observed optical, UV and X-ray continuum spectra.  The model is available for use with the {\sc xspec} spectral fitting package, under the name {\sc optxagnf}.

The model aims to capture the essential features of the continuum emission from a relativistic thin accretion disk with a Comptonizing corona in the equatorial plane of a rotating black hole.  It is assumed that material is accreted only through the outer disk, and the released gravitational energy is divided between three spectral components:
\begin{itemize}
	\item A pseudo-thermal accretion disk component
	\item A high-energy, power-law non-thermal component, formed by Compton up-scattering from an optically thin, high-temperature medium
	\item An intermediate (`soft X-ray excess') component formed by Compton up-scattering from a Compton-thick, low-temperature medium
\end{itemize}

The model implements the following:
\begin{itemize}
\item The standard \citet{Novikov_Thorne_1973} thin-disk relativistic accretion disk emissivity 
\item A disk spectrum produced by summing blackbodies in successive annuli, with the temperature of the blackbody spectrum corrected if necessary using an empirical color temperature correction $f_{\rm{col}}$.  The correction is required due to the absorption opacity varying with temperature, density and wavelength; in particular it is used where hydrogen is ionized in the inner parts of the disk in an attempt to account for modifications to the spectrum due to the disk not being locally thermalized and due to electron-scattering opacity \citep{Davis_Done_Blaes_2006}.  The correction becomes increasingly important as the temperature of the disk increases (as the luminosity increases or black hole mass decreases).
\item The disk luminosity between the innermost stable circular orbit $r_{\rm{ISCO}}$ and $r_{\rm{cor}}$, the `coronal radius', is all emitted as Comptonized radiation.
\item The Comptonized radiation is split between the soft, intermediate component (assumed to arise in the accretion structure itself) and the hard X-ray power-law tail.  The seed photons for Comptonization in both cases come from a blackbody with the same (corrected) temperature as the disk at $r_{\rm{cor}}$.
\item The total luminosity from all three components of the emission (disk, Comptonized soft X-ray excess, Comptonized hard X-ray tail) is fixed by the accretion rate and SMBH spin.
\end{itemize}

The following are the parameters required by {\sc optxagnf} to characterize the various physical components of the model:
\begin{itemize}
\item $M_\mathrm{BH}$, the black hole mass
\item $L/L_{\textrm{Edd}}$, the AGN luminosity in units of the Eddington luminosity, which is proportional to $M_\mathrm{BH}$.
\item $a^*$, the dimensionless black hole rotation parameter, between 0 and 1 (assuming prograde accretion)
\item $r_{\rm{cor}}$, the coronal radius (inner edge of the visible portion of the disk), in units of the gravitational radius ($r_g = GM_\mathrm{BH}/c^2$, proportional to $M_\mathrm{BH}$)
\item $r_{\rm{out}}$, the outer radius of the modeled disk, in units of the gravitational radius, proportional to $M_\mathrm{BH}$
\item $T$, the temperature of the Compton optically thick material which produces the soft X-ray excess
\item $\tau$, the optical thickness of the Compton optically thick material
\item $f_\mathrm{PL}$, the proportion of the corona power emitted in the hard power-law component; $1 - f_\mathrm{PL}$ is the proportion in the Compton optically thick component
\item $\Gamma$, the negative of the photon power-law index of the hard X-ray tail component
\end{itemize}

\subsection{Relationships between {\sc optxagnf} model parameters}
\label{sec:optxagnf_degeneracies}
We explored the behavior of the disk and non-thermal power-law components of the {\sc optxagnf} model across a wide parameter space (the parameter ranges given in Section~\ref{sec:reparametrizing}) and showed that there are strong spectral-shape degeneracies between key parameters.  The following sections discuss these degeneracies, which are in addition to the total degeneracy between $M_\mathrm{BH}$ and $L/L_{\textrm{Edd}}$ which originates in the Novikov-Thorne model.

\subsubsection{Relationship between disk coronal radius, SMBH mass and Eddington rate}
Higher disk temperatures lead to higher-energy disk emission, and in particular the high-energy cutoff of the disk emission is set by the maximum temperature of the directly visible part of the accretion disk.  Increasing the temperature of the highest-temperature visible parts of the disk and thereby producing higher-energy disk emission is achievable by increasing $(L/L_{\mathrm{Edd}})/M_\mathrm{BH}$ (increasing the luminosity and/or decreasing the mass), or alternatively by reducing $r_{\rm{cor}}$.  Consequently $r_{\rm{cor}}$ is partially degenerate in its effects on the disk spectrum with $L/L_{\mathrm{Edd}}$ and $M_\mathrm{BH}$.  The degeneracy is not total, because changing the innermost visible normalized radius modifies the shape of the disk spectrum in ways that changing $(L/L_{\mathrm{Edd}})/M_\mathrm{BH}$ cannot.

\subsubsection{Relationship between SMBH spin and disk coronal radius}
\begin{figure}
  \centering
      \includegraphics[width=0.49\textwidth]{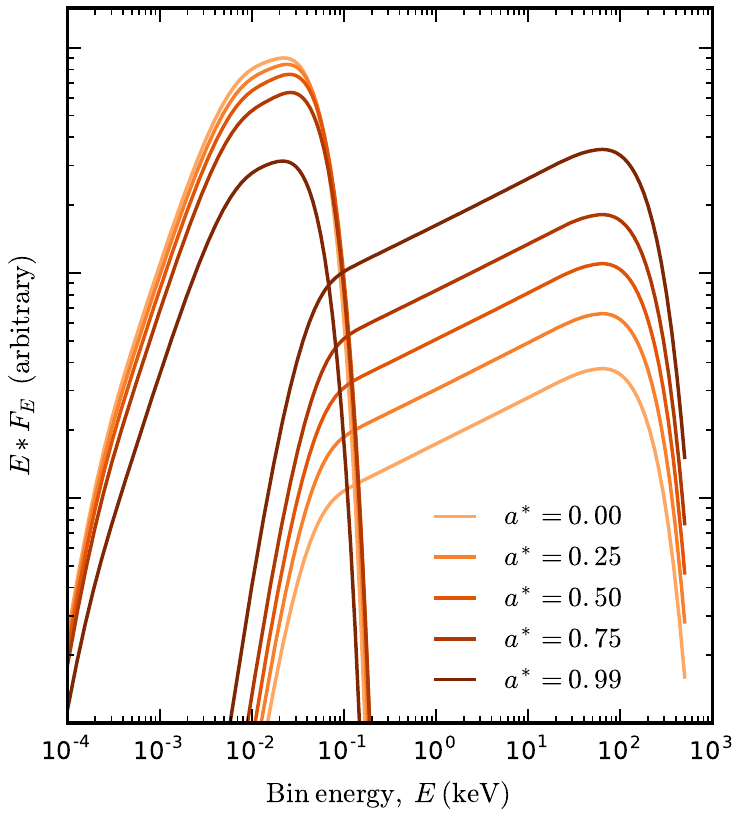}
  \caption{The disk and high-energy non-thermal emission components for varying spin parameter $a^*$, with the other key {\sc optxagnf} parameters fixed at $M_\mathrm{BH} = 10^8 \; M_\odot$, $L/L_{\mathrm{Edd}} = 0.1$, $r_{\rm{cor}} = 10 \; r_g$, and $\Gamma = 1.8$.  The intermediate component is not considered ($f_\mathrm{PL} = 1$).  Varying $a^*$ changes the relative contributions of the disk and non-thermal components to the total spectrum.  Note that the shape of these components remains relatively independent of $a^*$.}
  \label{fig:a_vary}
\end{figure}

Changing the spin $a^*$ with the other parameters fixed has the effect of changing the relative flux contribution of the disk and non-thermal components.  This effect is illustrated in Figure~\ref{fig:a_vary}, which shows models with no intermediate component.  As the spin of a SMBH increases, more energy is released by infalling matter; the additional energy is released close to the SMBH event horizon.  When we fix the mass, coronal radius and luminosity, increasing the spin serves primarily to apportion more of the fixed luminosity to the part of the disk within the coronal radius, and hence to the non-thermal component.   Hence $a^*$ has a similar effect to $r_{\rm{cor}}$ in that it changes the relative contributions of the disk and non-thermal components.

Figure~\ref{fig:a_vary} shows that the shape of the two spectral components is not strongly affected by $a^*$ (with $r_{\rm{cor}} = 10$~$r_g)$. However the parameters $r_{\rm{cor}}$ and $a^*$ are not entirely degenerate in their effects on the spectral shape, since $a^*$ affects the spectral shape through modifying the radial functions $\mathcal{B}(r, a^*)$, $\mathcal{C}(r, a^*)$ and $\mathcal{Q}(r, a^*)$ in Equation~\ref{eq:T1}, whereas $r_{\rm{cor}}$ determines which values of $r$ contribute to the disk emission.

The full and partial degeneracies discussed in the previous sections suggest that $M_\mathrm{BH}$, $L/L_{\mathrm{Edd}}$, $r_{\rm{cor}}$ and $a^*$ could be combined into two parameters in a simplified model of the spectral shape - one parameter which shifts the disk emission in energy space, and one parameter which controls the relative flux between the disk and non-thermal components.

\section{Shrinking the spectral model parameter space}
\label{sec:param_reduction}

We reduce the free parameter space of the {\sc optxagnf} model by fixing or neglecting some parameters, and by reparametrizing the remaining parameter space.

\subsection{Neglected or fixed parameters}
\label{sec:neg_params}

\subsubsection{Intermediate soft X-ray component}
\label{sec:ignoring_soft_excess}
The origin of the widely-observed soft X-ray excess in AGN remains uncertain, despite recent progress.  In particular, two general explanations are currently favored, which attempt to explain the relatively constant energy of the soft excess ($kT_e$~${\sim}0.1-0.2$~keV).  These explanations are Comptonized disk emission (e.g.\ the treatment in {\sc optxagnf}), and blurred reflection from the ionized accretion disk \citep[e.g.][]{Walton_2013_reflection}.  Additionally, absorption has been seriously considered \citep[e.g.][]{Gierliski_Done_2004}, and there is evidence that a contribution to the soft X-ray excess comes from the wider NLR \citep[e.g.][]{Bianchi_2006_xray_excess}.

The possible physical processes that give rise to the centrally-originating soft excess may be constrained using hard and soft time lags (hard X-rays lagging soft X-rays and soft X-rays lagging hard X-rays) and correlations between spectral components.  The time-delay of soft X-ray time lags is strongly correlated with black hole mass \citep{De_Marco_2013_AGN_X-ray_lags}.


Each soft X-ray excess photon with energy in the range ${\sim}0.2-2$~keV will have less effect on the ionization state of the nebula than a photon with energy in the range $0.01-{\sim}0.2$~keV.  In the pilot study of \citet{Dopita_2014_S7_I} we found that the intermediate soft X-ray component was not required to achieve satisfactory fits to the optical emission-line spectrum of the Seyfert galaxy NGC~5427 with photoionization models, although the observed spectrum did not include high-ionization lines.  We do however expect that the abundances of high ionization potential species such as \ion{Ar}{10} (423~eV), \ion{Fe}{14} (361~eV) and \ion{Fe}{10} (234~eV) would be affected by dominant soft X-ray excesses.  Additionally, a strong soft X-ray excess extends the partially-ionized zone, where Auger electrons produced by X-ray bombardment partially ionize the gas, and therefore must to some extent enhance emission of lines such as [\ion{O}{1}] and [\ion{S}{2}].

\begin{figure*}
	\centering
	\includegraphics[width=1.0\textwidth]{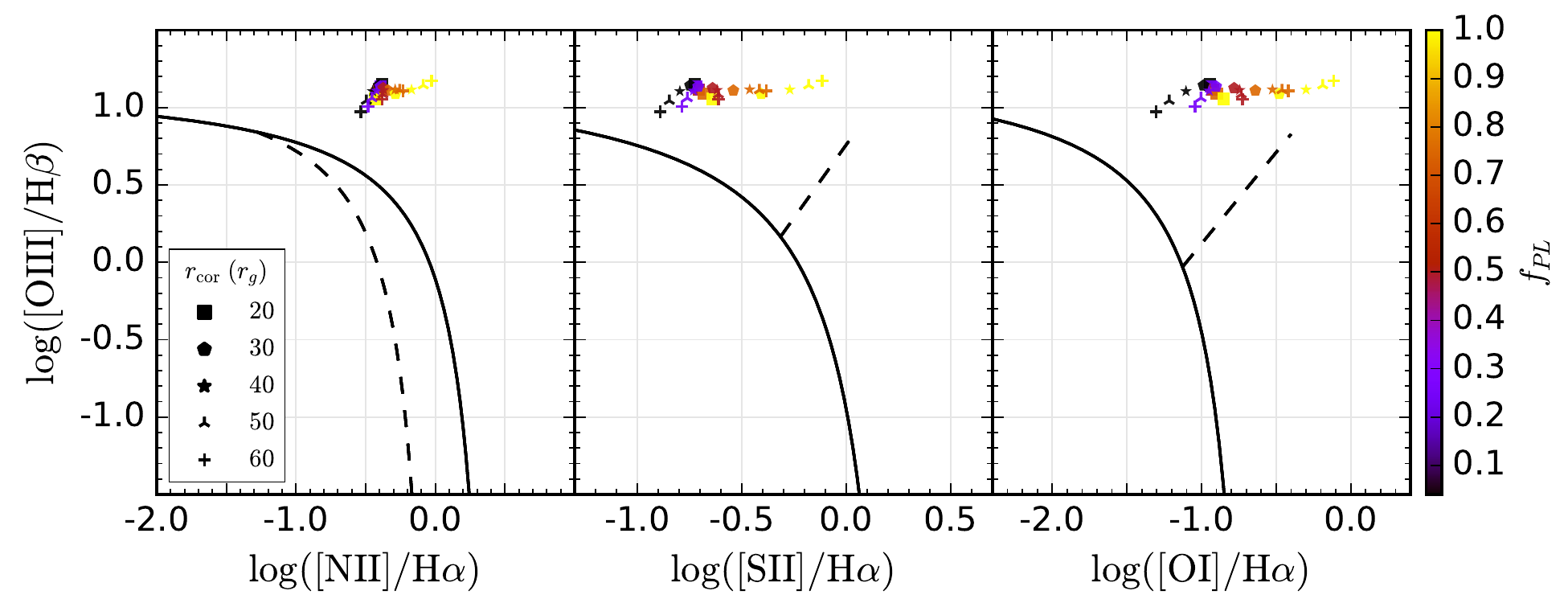}
	\caption{Predicted emission-line ratios produced by {\sc mappings} photoionization models using {\sc optxagnf} input ionizing spectra, shown on optical diagnostic diagrams \citep{BPT_1981, 1987VO}.  The solid lines \citep{Kewley_2001_starburst} and the dashed line in the left panel \citep{Kauffmann_2003_AGN} divide \ion{H}{2} regions (below) and nebulae with higher ionization states (above).  The dashed lines in the rightmost panels \citep{Kewley_2006_AGN_hosts} divide LINERs (below) and Seyferts (above).  The fraction of the energy inside the coronal radius that goes to the power-law component as opposed to the intermediate Comptonized component was set to each of $f_{PL} =$~0.04, 0.25, 0.5, 0.75, 1.00 and for each of these values we considered coronal radii of $r_\mathrm{cor} =$~20, 30, 40, 50, 60~$r_g$.  Points are colored by $f_{PL}$ and the marker shape indicates $r_\mathrm{cor}$.  We argue in Section~\ref{sec:ignoring_soft_excess} that the results in this figure demonstrate that photoionization modeling of the strong optical lines is insensitive to the soft X-ray excess.  For this experiment the {\sc mappings} models were configured with metallicity $Z = 2$~$Z_\odot$, ionization parameter $U(H) = 10^{-2}$, and a constant pressure of $P/k = 10^6$~K$\,$cm$^{-3}$.  The other parameters of {\sc optxagnf} were set to $M_\mathrm{BH} = 5 \times 10^7$~$M_\odot$, $L/L_{\mathrm{Edd}} = 10^{-0.5}$, $a^* = 0$, $\Gamma = 2.0$, and (for the Comptonized soft excess) $kT = 0.2$~keV with $\tau = 15$.}
	\label{fig:BPTVO_soft_excess}
\end{figure*}

We performed an investigation on the effect of the soft excess on photoionization modeling of NLR clouds, using ionizing spectra from the {\sc optxagnf} model.  The parameters of the {\sc mappings}~5.1\footnote{Available at \url{http://miocene.anu.edu.au/Mappings}} (Sutherland et al. 2016, in prep.) models and {\sc optxangf} spectra were fixed, apart from two quantities: the relative strength of the soft excess which ranged from $0\%$ to $100\%$ of the non-thermal flux, and the coronal radius which ranged from $r_\mathrm{cor} = 20$ to $r_\mathrm{cor} = 60$.  The models were ionization-bounded and ended when the nebula was 99\% neutral.  The results are presented in standard optical diagnostic diagrams \citep{BPT_1981, 1987VO} in Figure~\ref{fig:BPTVO_soft_excess}, with the fiducial model parameter values listed in the caption.  The standard optical diagnostic diagrams use ratios of lines associated with different ionization levels in the gas to distinguish nebulae based on the cause of ionization.  Ionization increases from bottom-left to top-right on the diagrams.  The lower-left of each of the diagnostic diagrams is associated with \ion{H}{2} regions of varying metallicity and ionization parameter, whereas the top-right is associated with ionization by harder AGN spectra or shocks.  The standard dividing lines on the diagrams are described in the caption.

The results shown in Figure~\ref{fig:BPTVO_soft_excess} confirm that the soft X-ray excess is not of primary importance when predicting diagnostic optical emission-line ratios.  The line ratios are evidently most sensitive to variation in the coronal radius or weighting of spectral components when most of the energy released inside the coronal radius is emitted in the power-law component as opposed to the soft excess.  For the tested $r_\mathrm{cor}$ values larger than 20~$r_g$, the [\ion{S}{2}]/H$\alpha$ and [\ion{O}{1}]/H$\alpha$ ratios are significantly increased by increasing the proportion of the flux emitted in the power-law tail, resulting in some outlying points to the right of the second and third panel in the figure.  The positions of these points are due to the partially-ionized zone being extended by hard X-rays in improbably hard ionizing spectra.  If we do not consider these power-law dominated models (shown for completeness), the range of points representing the different soft excess strengths is small compared to the variation that may be produced by varying the ionization parameter or metallicity, for example.  If the proportion of the energy emitted in the soft excess as opposed to the power-law component does not vary as strongly between real objects as it does in our experiment, and additionally if real objects are mass-bounded as opposed to ionization-bounded (i.e.\ hard X-rays escape the NLR), then the ranges shown in Figure~\ref{fig:BPTVO_soft_excess} would represent approximate upper bounds on the sensitivity to the soft excess.

In the remainder of this work we do not include the intermediate Comptonized component, such that the fraction of the corona power in the power-law tail is set to $f_{PL}=1$, and the parameters for the temperature and optical thickness of the Comptonizing optically-thick medium have no effect.  

Neglecting the intermediate component enables a vast simplification of the model.  Including the soft excess would complicate the comparison of photoionization models to observed NLR emission-line spectra due to uncertainty regarding the origin, time-averaged behavior, and importance relative to other spectral components of the soft X-ray excess.  We note that recently \citet{Pal_2016_X-ray_AGN} attributed ${\sim}60-80\%$ of the ${\sim}0.3-2$~keV emission observed in the AGN II~Zw~177 to the soft excess, with significant variation observed between observations separated by 11~years.  Another recent work is a study of the AGN RE~J1034+396 \citep{Czerny_2016}, in which the best-fitting models suggested that ${\sim}90\%$ of the power produced inside the coronal radius is emitted in the intermediate component.  Although the strength of the soft excess has been measured in sources such as II~Zw~177 and RE~J1034+396, the properties of the soft X-ray excess across the population of Seyfert galaxies, not just those that have been studied in X-rays, are unknown.

\subsubsection{The black hole spin}
\label{sec:BH_spin_neglected}

The spins of astrophysical black holes have long been expected to be non-zero.  It is now routine to `map' the central ${\sim} 20$~$r_g$ of AGN accretion structures by studying X-ray reverberation lags and blurred accretion disk reflection \citep{Fabian_2016_BH_accretion}.  Identifying the innermost radius required by spectral blurring analysis with the Innermost Stable Circular Orbit (ISCO) allows the black hole spin to be measured.  Spin measurements performed to date generally demonstrate high spins \citep{Reynolds_2014_BH_spin}.  However because radiative efficiency increases by up to a factor of 5 with increasing spin, the very high calculated spins for many AGN with spin measurements are associated with a strong selection bias \citep{Fabian_2016_BH_accretion}.

As demonstrated above in Section~\ref{sec:optxagnf_degeneracies}, the black hole spin is partially degenerate with $r_{\rm{cor}}$ in that it changes the relative contributions of the disk and non-thermal components to the total spectrum.  The coronal radius will be set larger than radii for which we expect the disk spectrum shape to be strongly affected by the SMBH spin.  Figure~\ref{fig:a_vary} shows that changing the spin changes the relative energy in the disk and non-thermal components by an order of magnitude (between spins of 0 and 1); in our simplified model this effect will be accounted for by an explicit parameter for this scaling.  The SMBH spin was set to $a^* = 0$.

\subsubsection{Outer edge of accretion disk}
The outer disk produces non-ionizing optical and IR emission, and therefore does not contribute to the NLR heating.  The parameter defining the outer edge of the accretion disk was set to $10^4$~$r_g$, a default value.

\subsection{Reparametrizing the AGN continuum model}
\label{sec:reparametrizing}

As a result of the simplifications discussed in the Section~\ref{sec:neg_params}, the {\sc optxagnf} parameters have been reduced to the four most important parameters $M_\mathrm{BH}$, $L$, $r_{\rm{cor}}$ and $\Gamma$.  Exploration of the properties of the theoretical AGN spectra and development of a simplified model (Section~\ref{sec:oxaf}) were performed over the following ranges of these four parameters:
\begin{itemize}
\item $-5.0 \leq \log_{10}(L/L_{\mathrm{Edd}}) \leq 0.0$
\item $6.0 \leq \log_{10}(M_\mathrm{BH}/M_\odot) \leq 9.0$
\item $1.4 \leq \Gamma \leq 2.6$
\item $10$~$r_g \leq r_{\rm{cor}} \leq 100$~$r_g$
\end{itemize}

The maximum luminosity of Eddington was chosen because the Novikov-Thorne thin disk model cannot plausibly be used beyond this limit; indeed a \textit{slim} disk prescription is necessary for a luminosity above ${\sim}0.3$~$L/L_{\mathrm{Edd}}$.  The range of SMBH masses covers the observed range in AGN except for extreme objects (c.f.\ Figure~4 in \citet{2009_Vika_SMBH_M_func}).  The broad range in $\Gamma$ covers the approximate range of observed values (c.f.\ Figure~4 in \citet{2013_IBIS_Xray_AGN}), corresponding to very soft ($\Gamma$ = 2.6) through to very hard ($\Gamma$ = 1.4) power-law tails.

The physical extent of the corona has been constrained using analysis of blurred X-ray reflection, modeling of X-ray reverberation, and microlensing observations; these methods suggest that the Comptonizing corona lies within approximately $10$~$r_g$ of the black hole \citep{Fabian_2016_BH_accretion}.  However because of the selection bias discussed in Section~\ref{sec:BH_spin_neglected}, coronal geometries inferred in X-ray studies are not necessarily representative of the population of Seyfert galaxies.  We chose a lower bound for the coronal radius of 10~$r_g$ here; our experimentation showed that below this value the high-energy part of the accretion disk emission becomes too strongly affected by relativistic corrections for easy simplification of the model.  Setting a minimum coronal radius of 10~$r_g$ will primarily affect the high-energy part of the accretion disk emission by removing it from the disk spectrum.  The top end of the $r_{\rm{cor}}$ range was selected to include the range suggested by \citet{Done_2012_AGN_SED}.

\subsubsection{Reparametrizing the disk component}

A highly desirable simplification is combining the three fully or partially degenerate parameters $L$, $M_\mathrm{BH}$ and $r_{\rm{cor}}$ into a single parameter that is able to parametrize (at least approximately) the position in energy space and shape of the Big Blue Bump (BBB) disk emission.  The parameter which was chosen to replace the degenerate trio ($L$, $M_\mathrm{BH}$, $r_{\rm{cor}}$) was $E_{\mathrm{peak}}$, the peak of the BBB when the disk spectrum is given in a log-log plot of energy flux versus energy (e.g.\ Figure~\ref{fig:a_vary}).  The location of the single peak is a reliable and easily-calculated feature of a model BBB.

\begin{figure}
  \centering
      \includegraphics[width=0.49\textwidth]{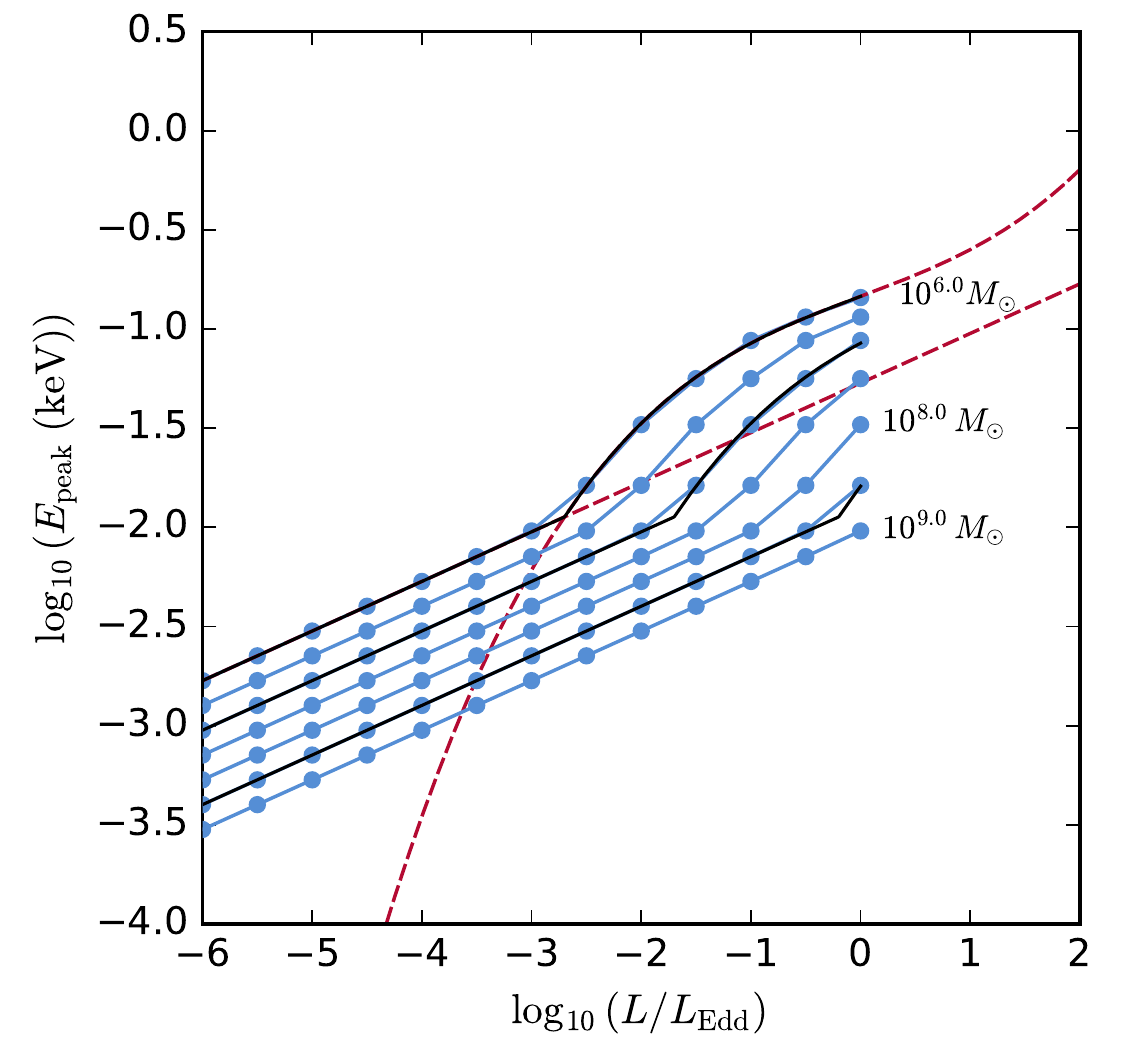}
  \caption{We can characterize the dominant AGN emission in the EUV, the Big Blue Bump (BBB), in terms of the peak of its energy distribution, $E_\mathrm{peak}$. Here we show the $E_\mathrm{peak}$ predicted by {\sc optxagnf} as a function of $L/L_\mathrm{Edd}$ (blue points) and $M_\mathrm{BH}$ (blue curves as labeled, spaced at 0.5 dex), for $r_\mathrm{cor} = 24.5$~$r_g$.  To predict $E_\mathrm{peak}$ for a given $r_\mathrm{cor}$ and $M_\mathrm{BH}$ we use a model with the functional form $\max(f_1(L/L_\mathrm{Edd}), f_2(L/L_\mathrm{Edd}))$ (Equation~\ref{eq_E_peak}) where $f_1$ is a power-law (at low $L/L_\mathrm{Edd}$) and $f_2$ is a log-space cubic (at high $L/L_\mathrm{Edd}$).  Both $f_1$ and $f_2$ are illustrated for $M_\mathrm{BH} = 10^6$ by the red dashed curves. Model fits for $M_\mathrm{BH} = 10^{6.0}, 10^{7.0}$ and $10^{8.5}$~$M_\odot$ are overlaid in black.
  }
  \label{fig:E_peak_pred_r24}
\end{figure}

The value of $E_{\mathrm{peak}}$ was determined for a grid of {\sc optxagnf} spectra over the relevant ranges of the three parameters affecting $E_{\mathrm{peak}}$, which are $L$, $M_\mathrm{BH}$ and $r_{\rm{cor}}$.  The results were systematically analyzed for each value of $r_{\rm{cor}}$, before all the analyses for various $r_{\rm{cor}}$ values were combined into an empirical formula for calculating $E_{\mathrm{peak}}$ as a function of ($L$, $M_\mathrm{BH}$, $r_{\rm{cor}}$).  An example of the analysis that was applied for a single value of $r_{\rm{cor}}$ is presented in Figure~\ref{fig:E_peak_pred_r24}.

Figure~\ref{fig:E_peak_pred_r24} demonstrates how fits for $E_{\mathrm{peak}}$ at a given $r_{\rm{cor}}$ must account for the `piecewise' behavior of $E_{\mathrm{peak}}$ as a function of $M_\mathrm{BH}$ and $L$.  Our solution to this problem was to use a combination of linear and cubic functions as described in the figure and caption.  The coefficients of these linear and cubic functions varied as a function of $r_{\rm{cor}}$.  To construct a general function applicable over the entire three-dimensional ($L$, $M_\mathrm{BH}$, $r_{\rm{cor}}$) parameter space, we required a means to determine the coefficients of the linear and cubic functions from $r_{\rm{cor}}$.  Linear and quadratic functions of $\log r_{\rm{cor}}$ produced satisfactory fits to the values of these coefficients.

The analysis in Figure~\ref{fig:E_peak_pred_r24} was applied for $r_{\rm{cor}} = 6$, 10, 15, 24, 39, 63, 100~$r_g$.  The analysis for $r_{\rm{cor}} = 6$~$r_g$ (i.e. at the innermost stable circular orbit, such that the corona is effectively absent) was found to be generally inconsistent with the analyses for the larger $r_{\rm{cor}}$ values, presumably due to the effect of strong general relativistic corrections.  We thus did not use the analysis for $r_{\rm{cor}} = 6$~$r_g$.

The combined analysis for the various $r_{\rm{cor}}$ values culminated in the following equation for a hypersurface to predict $E_{\mathrm{peak}}$ as a function of $L$, $M_\mathrm{BH}$ and $r_{\rm{cor}}$:

\begin{equation}
\log_{10} \; E_{\mathrm{peak}}(L, M_\mathrm{BH}, r_{\rm{cor}}) = \max( f_1, f_2 ) \label{eq_E_peak} \\
\end{equation}
where
\begin{align}
f_1(L, M_\mathrm{BH}, r_{\rm{cor}}) = (&A_{x^2} \, R(r_{\rm{cor}})^2 + A_{x^1} \, R(r_{\rm{cor}}) + A_{x^0}) \nonumber \\
       & + B \; S(L,M_\mathrm{BH})  \label{eq:f1}
\end{align}
and
\begin{align}
f_2(&L,M_\mathrm{BH}, r_{\rm{cor}}) = \nonumber \\
    & (b_{3,x^1} \, R(r_{\rm{cor}}) + b_{3,x^0}) (S(L,M_\mathrm{BH})+6)^3 \; + \nonumber \\
    & (b_{1,x^2} \, R(r_{\rm{cor}})^2 + b_{1,x^1} \, R(r_{\rm{cor}}) + b_{1,x^0}) (S(L,M_\mathrm{BH})+6) \nonumber \\
    & + \; (a_{x^2} \, R(r_{\rm{cor}})^2 + a_{x^1} \, R(r_{\rm{cor}}) + a_{x^0})
    \label{eq:f2}
\end{align}

Here the energy of the BBB peak $E_{\mathrm{peak}}$ is in keV, $S(L,M_\mathrm{BH}) = \log_{10}((L/L_{\rm{Edd}})/M_\mathrm{BH})$ where $M_\mathrm{BH}$ is in M$_{\odot}$, and $R(r_{\rm{cor}}) = \log_{10} (r_{\rm{cor}}/r_g)$.  The fit parameters are given in Table~\ref{tab:E_peak_params}.  Note that $L$ and $M_\mathrm{BH}$ appear explicitly in Equations~\ref{eq:f1} and~\ref{eq:f2} only through the quantity $S(L,M_\mathrm{BH})$, due to the complete degeneracy in their effects on the spectral shape (Section~\ref{sec:M_L_degeneracy}; note that there is an implicit dependence on $M_\mathrm{BH}$ in the normalization of $L$ and $r_{\rm{cor}}$).  The arbitrary constant 6 occurs with $S(L,M_\mathrm{BH})$ in Equation~\ref{eq:f2} only because it was propagated through the analysis from the cubic functions such as those in Figure~\ref{fig:E_peak_pred_r24}, where the cubic was fitted to the $\log M_\mathrm{BH}/M_\odot = 6$ data (the lowest $M_\mathrm{BH}$ value we considered), and was shifted appropriately to apply to other $M_\mathrm{BH}$ values. Equations~\ref{eq_E_peak} to~\ref{eq:f2} in conjunction with Table~\ref{tab:E_peak_params} predict $E_{\mathrm{peak}}$ with a standard deviation of ${\sim}2\%$ in linear energy space.

The shape of the BBB spectrum depends (to first order) only on the energies it covers, so we use $E_{\mathrm{peak}}$ to parametrize both the location in energy space and shape of the BBB emission.  The construction of a simplified BBB model using this approach is discussed in Section~\ref{sec:BBB_modeling}.

\begin{deluxetable}{rrrrrrrr}
\tablecolumns{4}
\tablewidth{0pc}
\tablecaption{Fit parameters for Equations~\ref{eq:f1} and \ref{eq:f2} \label{tab:E_peak_params}}
\tablehead{
\colhead{Parameter} & \colhead{Value} & \colhead{} & \colhead{Parameter} & \colhead{Value} }
\startdata
 $A_{x^2}$   & -0.18 $\pm$ 0.04   && $b_{1,x^2}$ &  0.391 $\pm$ 0.008  \\
 $A_{x^1}$   &  0.0 $\pm$ 0.1     && $b_{1,x^1}$ & -0.83 $\pm$ 0.02  \\
 $A_{x^0}$   &  0.59 $\pm$ 0.08   && $b_{1,x^0}$ &  0.61 $\pm$ 0.02  \\
 $B$         &  0.250 $\pm$ 0.004 && $a_{x^2}$   & -0.30 $\pm$ 0.03  \\
 $b_{3,x^1}$ &  0.034 $\pm$ 0.002 && $a_{x^1}$   &  0.4 $\pm$ 0.1  \\
 $b_{3,x^0}$ & -0.019 $\pm$ 0.003 && $a_{x^0}$   & -0.82 $\pm$ 0.07
\enddata
\end{deluxetable}

\subsubsection{Reparametrizing the non-thermal component}
 
The shape of the non-thermal component is determined by the shape of the seed photon spectrum and by the parameter $\Gamma$.  The seed photon spectrum is a blackbody with a temperature which will be determined by the single parameter $E_{\mathrm{peak}}$, so we have already reparametrized the shape of the non-thermal component such that it depends on only two parameters.  The construction of the non-thermal emission component is discussed in Section~\ref{sec:NT_modeling}.

\section{The {\sc oxaf} model}
\label{sec:oxaf}

This section describes the construction of the {\sc oxaf} model.  In particular {\sc oxaf} is a 3-parameter model, and the name {\sc oxaf}, a contraction of {\sc optxagnf}, was chosen to reflect that {\sc oxaf} is in some sense a `reduced' {\sc optxagnf}.

The three parameters of the {\sc oxaf} model are the energy of the peak of the Big Blue Bump (BBB) $E_{\mathrm{peak}}$, the photon power-law index $\Gamma$, and the proportion of the total energy flux emitted in the non-thermal component $p_\mathrm{NT}$, with the remainder being in the BBB.

In {\sc optxagnf} the shape of the BBB and the proportion of energy which goes to the power-law component are not independent, because both are affected by $r_{\rm{cor}}$.  However in {\sc oxaf}, the parameters $E_{\mathrm{peak}}$ (which determines the BBB shape) and $p_\mathrm{NT}$ are independent by construction.  For a given $E_{\mathrm{peak}}$, {\sc oxaf} does not have any parameters available to reproduce the variation in {\sc optxagnf} BBB shapes which occurs as $r_{\rm{cor}}$ changes.  However the variation of the BBB shape with $r_{\rm{cor}}$ is a small effect, so a single {\sc oxaf} BBB is able to satisfactorily reproduce {\sc optxagnf} BBBs which have the same peak energy but a range of $r_{\rm{cor}}$ (Section~\ref{sec:comparison}).

The following sections show how we construct a BBB using only one parameter, the energy of the BBB peak $E_{\mathrm{peak}}$, and a non-thermal component using only two parameters ($E_{\mathrm{peak}}$ and $\Gamma$).

\subsection{Modeling the accretion disk emission}
\label{sec:BBB_modeling}

The BBB is parametrized using only one parameter, the energy of the BBB peak $E_{\mathrm{peak}}$.  The {\sc oxaf} model must use this input to determine the shape of the BBB, in particular for relatively high BBB peak energies, when the color correction becomes important.

\begin{figure}
  \centering
      \includegraphics[width=0.47\textwidth]{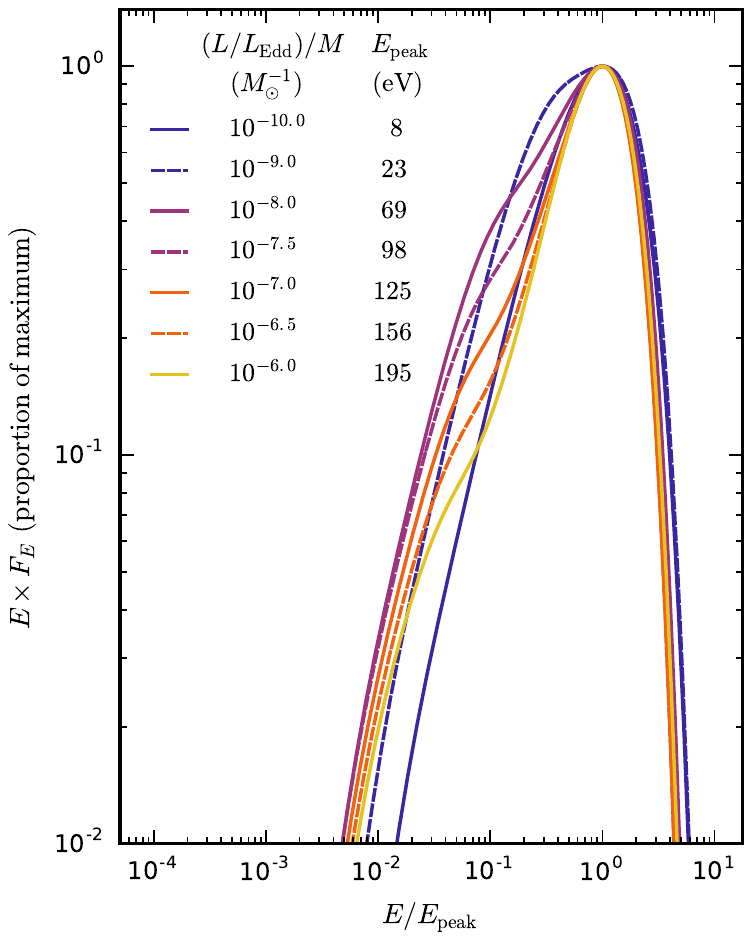}
  \caption{Normalized Big Blue Bump (BBB) accretion disk component {\sc optxagnf} spectra, shown for varying $(L/L_{\mathrm{Edd}})/M_\mathrm{BH}$ and for $r_{\mathrm{cor}} = 10$~$r_g$.  The models were generated using a standalone version of {\sc optxagnf} produced using a simple fortran wrapper.  The shape of the BBB changes due to the increasingly important color correction for higher-energy BBBs.  The curves are normalized to have the same flux in the bin with maximum flux, and shifted in logarithmic energy space to have coinciding peaks.}
  \label{fig:BBB_shape}
\end{figure}

The modeled BBBs are considered in terms of log-log plots of energy flux (i.e.\ keV/cm$^2$/s) versus energy.  The effects of the color correction in this space are illustrated in Figure~\ref{fig:BBB_shape}.

The {\sc oxaf} model constructs the BBB as follows: sixth-order polynomials were fit to a series of {\sc optxangf} BBBs at different energies ($\log_{10}(E_{\mathrm{peak}}/\mathrm{keV})$~=~$-3.0$, $-2.4$, $-1.9$, $-1.4$, $-1.2$, $-0.7$) roughly representative of the range of shapes shown in Figure~\ref{fig:BBB_shape}.  The BBBs were normalized to have the same peak flux and position of the peak flux in energy space before fitting.  The coefficients of the fits are stored in the {\sc oxaf} code.  To determine the shape of an {\sc oxaf} BBB for a given input $E_{\mathrm{peak}}$, the code (log-)linearly interpolates between the polynomial fits for the nearest $E_{\mathrm{peak}}$ values above and below the input $E_{\mathrm{peak}}$.  The resulting polynomial BBB shape is a weighted sum of the two `nearest' polynomials.  The interpolated polynomial is shifted in energy space so the peak aligns with the input $E_{\mathrm{peak}}$ value.

\subsection{Modeling the non-thermal emission}
\label{sec:NT_modeling}
The non-thermal component is parametrized using two quantities: $E_{\mathrm{peak}}$ and $\Gamma$.

The {\sc oxaf} model uses the {\it Nthcomp} routine to generate an inverse-Compton scattered non-thermal component.  The fortran subroutines \textit{donthcomp} \citep{Zdziarski_1996, Zycki_1999}, and \textit{thcompton} with \textit{thermlc} \citep{Lightman_Zdziarski_1987} from {\sc xspec} were incorporated into {\sc oxaf}.  

There is one difference in the modeling of the non-thermal component between {\sc oxaf} and {\sc optxagnf}.  In {\sc optxagnf} the seed blackbody spectrum temperature is set to $T_\mathrm{cor}$, the color-corrected temperature of the disc at $r_{\mathrm{cor}}$.  To approximate this method in {\sc oxaf} it was necessary to convert the $E_{\mathrm{peak}}$ parameter to an estimate of $T_\mathrm{cor}$, despite the {\sc oxaf} model lacking an explicitly calculated $r_{\mathrm{cor}}$.  The conversion was achieved by extracting $T_\mathrm{cor}$ values for 72 representative {\sc optxagnf} models across the parameter space and relating them to the $E_{\mathrm{peak}}$ values calculated for these models using Equation~\ref{eq_E_peak}.  The following linear fit from this comparison is used by {\sc oxaf} to relate the two parameters:

\begin{equation}
\log_{10}(kT_\mathrm{cor}) = 0.96 \, \log_{10}(E_\mathrm{peak}) - 0.39
\label{eq:T_r_cor}
\end{equation}
where both $kT_\mathrm{cor}$ and $E_{\mathrm{peak}}$ are measured in keV.  Here the gradient and intercept have errors of 0.02 and 0.03~dex respectively, and the resulting error on $\log_{10}(kT_\mathrm{cor})$ is 0.07~dex.  The non-thermal spectrum is constructed by converting $E_{\mathrm{peak}}$ to $T_\mathrm{cor}$ using Equation~\ref{eq:T_r_cor} before inputting $\Gamma$, $T_\mathrm{cor}$ and an assumed Comptonizing plasma temperature of 100~keV into the {\sc oxaf} \textit{donthcomp} function, which in turn calls \textit{thcompton}, which calls \textit{thermlc}.

\subsection{Implementation}
\label{sec:implementation}
The {\sc oxaf} model was implemented as a module {\it oxaf.py} written in the programming language python, with a design focus on convenience of use.  The module depends only on the standard third-party numerical Python library {\it numpy}, is a single file which may be used with both python~2 and python~3, and requires no compilation step to run (the included fortran codes were converted to python).  The module may simply be run as a command line script to output a model spectrum to stdout, or alternatively may be imported to be used in python code.  The module is thoroughly self-documented and is available on the Astrophysics Source Code Library (ASCL) \footnote{Available at \url{https://github.com/ADThomas-astro/oxaf} (Not yet submitted to ASCL)}.

The {\it oxaf.py} module includes functions to find $E_{\mathrm{peak}}$, the peak of the Big Blue Bump (BBB) disk emission (i.e.\ implement Equation~\ref{eq_E_peak}), calculate an accretion disk spectrum (as described in Section~\ref{sec:BBB_modeling}), calculate a non-thermal component spectrum (as described in Section~\ref{sec:NT_modeling}), and sum these two components using a given weighting.

The output spectrum is a function of only three parameters, being $\log_{10}(E_{\mathrm{peak}}/\mathrm{keV})$, $\Gamma$, and $p_\mathrm{NT}$, which is the proportion of the total flux over the range $0.01 < E$~(keV)~$< 20$ which is assigned to the non-thermal component, with $1-p_\mathrm{NT}$ being the proportion assigned to the BBB disk component.

The functions which return the individual components and the full spectrum (sum of the two components) normalize the spectra so that the sum of the bin fluxes over the range $0.01 < E \mathrm{(keV)} < 20$ is equal to one.

\subsection{Validation}
\label{sec:comparison}
The {\sc oxaf} model was validated against the {\sc optxagnf} model by comparing {\sc oxaf} output with {\sc optxagnf} output for many models across the parameter space.  The BBB and non-thermal components were considered separately.  The {\sc optxagnf} parameters used were chosen so that their various combinations did not give multiple identical BBB peak energies and hence multiple identical {\sc oxaf} spectra.  The chosen values were $\log_{10}(L/L_{\textrm{Edd}})$~=~$-5.00$, -3.75, -2.85, -2.15, -1.60, -1.20, -0.80, -0.40, 0.00; $\log_{10}(M_\mathrm{BH}/M_\odot)$~=~$6.0$, 6.5, 7.0, 7.5, 8.0, 8.5, 9.0; $r_{\rm{cor}}$~=~$10$, 25, 40, 50, 63, 100~$r_g$ and $\Gamma$~=~1.4, 1.7, 2.0, 2.3, 2.6, for a total of 1890 {\sc optxagnf} spectra.  The range of parameters was chosen for completeness; not all of the tested parameter space necessarily corresponds to observed AGN.

Figures~\ref{fig:oxaf_validation_BBB} and \ref{fig:oxaf_validation_NT} in the appendix demonstrate the performance of the {\sc oxaf} model in reproducing the output spectra of {\sc optxagnf}.  The figures show that over the relevant energy range, from the Lyman limit to above the energy at which iron may be fully ionized (0.01 to ${\sim}$20~keV), {\sc oxaf} predictions are sufficiently close to {\sc optxagnf} predictions for our purposes (differences are far less than modeling uncertainties).  The interquartile range displayed in each figure indicates that over the relevant energy range, {\sc oxaf} predictions tend to be within ${\sim}$10-20\% of the {\sc optxagnf} value for the BBB component, and within ${\sim}5\%$ for the non-thermal component.  The largest deviations tend to occur for energies which are probably too high to have a strong impact on photoionization model results (1-10~keV) or are in an unlikely part of the parameter space (e.g.\ $r_{\rm{cor}} {\sim} 60$~$r_g$).  The most extreme discrepancies (i.e.\ 95th percentile and upwards) for the BBB comparisons tend to be due to irrelevant cases where only a small fraction of the total flux is in the relevant ionizing energy range, and hence the normalization over this range inflates a small section of the spectrum and magnifies proportional differences between the two models.  For the non-thermal component the differences are due only to the small scatter in the fit given in Equation~\ref{eq:T_r_cor}, i.e.\ the seed blackbody spectra for {\sc oxaf} and {\sc optxagnf} have slightly different temperatures.

The {\sc oxaf} spectra reproduce the {\sc optxagnf} models with sufficient accuracy for photoionization modeling of optical diagnostic emission lines.  The differences between {\sc oxaf} and {\sc optxagnf} are smaller than the uncertainties due to, for example, any of the following:
\begin{itemize}
\item The angular dependence of the accretion disk spectrum, including the effects of special relativistic beaming, gravitational beaming/light bending, and limb darkening \citep{Laor_Netzer_1989_accretion_disks}.  Gravitational redshifts are an additional consideration.
\item Uncertainties in fundamental aspects of the thin disk model, such as the $\alpha$-prescription, and aspects of the geometry of the accretion disk and corona including the variation with black hole spin \citep[e.g.][]{You_2012_GR_BH_spectrum}.  The luminosity range over which the model is applicable is another consideration, e.g.\ for luminosities over ${\sim}0.3$~$L_{\mathrm{Edd}}$ a `slim' disk is a more appropriate model.
\item Radiative transfer effects; for example, the use of a color correction in the {\sc optxagnf} model neglects H and He edges \citep[see Figure~1 in][]{Done_2012_AGN_SED}
\end{itemize}
Two simple tests were performed to compare {\sc mappings}~5.1 photoionization models based on {\sc optxagnf} and {\sc oxaf} spectra.  For the first test the input {\sc optxagnf} spectrum was configured with $(L/L_{\mathrm{Edd}},$~$M_\mathrm{BH}$,~$r_{\mathrm{cor}}$,~$\Gamma) =$ $(0.2,$~$10^7$~$M_\odot$,~$15$~$r_g$,~2.2), while the {\sc oxaf} spectrum was configured with $(E_{\mathrm{peak}},$~$\Gamma$,~$p_{\mathrm{NT}}) =$ (68.3~eV,~2.2,~0.223), where $E_{\mathrm{peak}}$ was calculated from Equation~\ref{eq_E_peak} and $p_{\mathrm{NT}}$ was calculated to match the proportion in the {\sc optxagnf} spectrum.  The plane-parallel {\sc mappings} models for this test were configured with a metallicity of $2 Z_\odot$, an ionization parameter of $U(H) = 10^{-2}$, and a constant pressure of $P/k = 10^6$~K$\,$cm$^{-3}$.  The resulting line fluxes were broadly consistent between the two models.  Of 43 optical lines with predicted fluxes of more than 1\% of H$\beta$, all had a flux discrepancy of less than 10\% of the flux predicted using the {\sc optxagnf} model.  All but 4 line fluxes showed a difference of less than 5\%; these 4 lines from \ion{N}{1} or \ion{O}{1} had differences of ${\sim}7-8\%$.  The reason for this discrepancy is that these lines from low-ionization species are very sensitive to small differences in the shape of the hard photon spectrum since the emission is produced mostly in a partially ionized zone heated by Auger electrons.

For the second test the input {\sc optxagnf} spectrum was configured with $(L/L_{\mathrm{Edd}},$~$M_\mathrm{BH}$,~$r_{\mathrm{cor}}$,~$\Gamma) =$ $(0.3,$~$10^8$~$M_\odot$,~$20$~$r_g$,~1.8), while the {\sc oxaf} spectrum was configured with $(E_{\mathrm{peak}},$~$\Gamma$,~$p_{\mathrm{NT}}) =$ (19.9~eV,~1.8,~0.342).  The {\sc mappings} models used a metallicity of $3 Z_\odot$, an ionization parameter of $U(H) = 10^{-2.5}$, and a constant pressure of $P/k = 10^6$~K$\,$cm$^{-3}$.  With the somewhat softer ionizing spectrum and different nebular parameters, there were only 31 optical lines with predicted fluxes of more than 1\% of H$\beta$, and all of those lines had a flux discrepancy below $2\%$, with 24 having a discrepancy under $1\%$.

The accuracy of {\sc oxaf} is more than sufficient for our purposes of photoionization modeling.

\section{The effect of {\sc oxaf} parameters on predicted emission-line ratios}
\label{sec:oxaf_predictions}

\begin{figure*}
	\centering
	\includegraphics[width=1.0\textwidth]{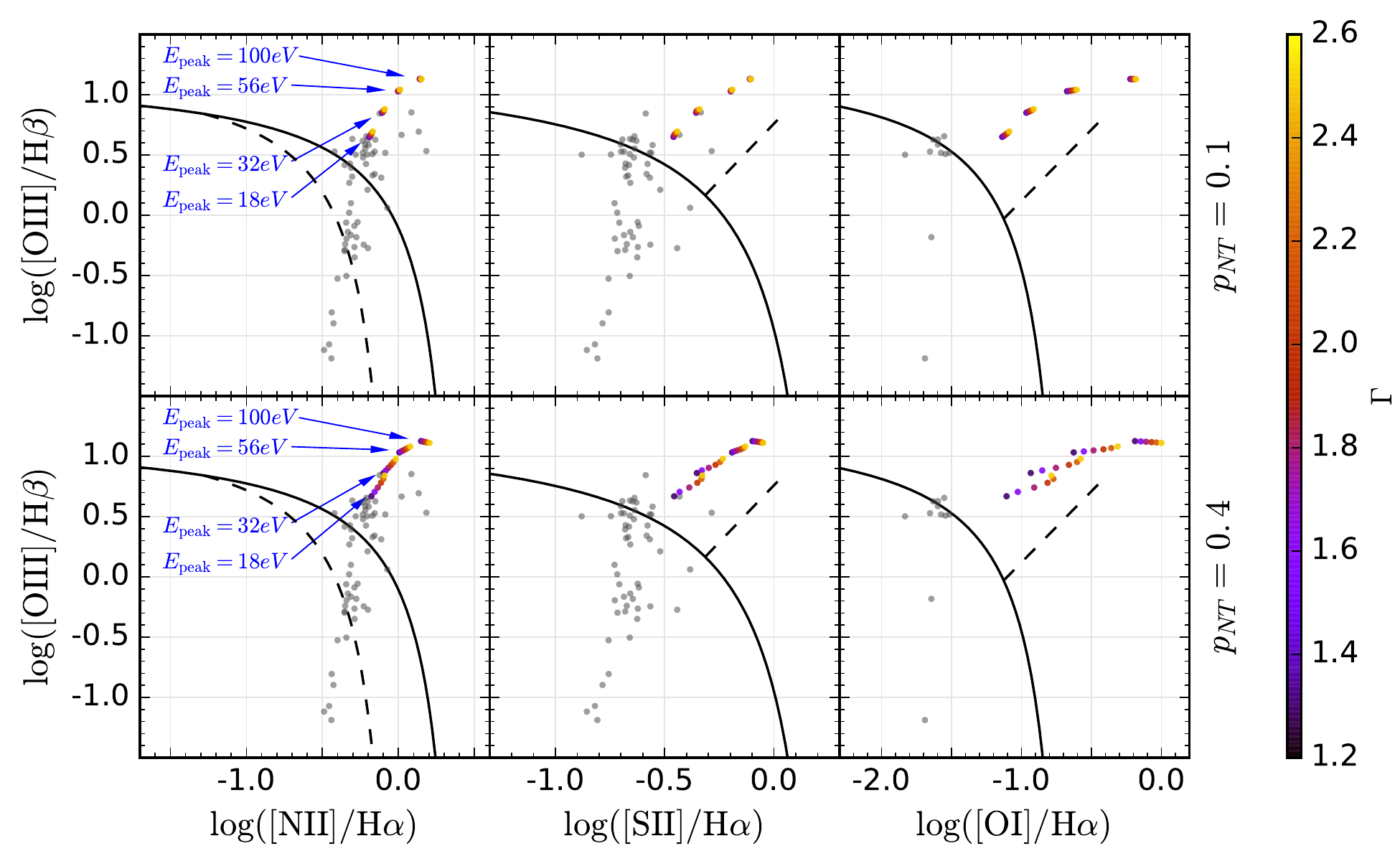}
	\caption{Optical diagnostic diagrams showing the effect of the {\sc oxaf} parameters $E_\mathrm{peak}$, $\Gamma$ and $p_\mathrm{NT}$ on {\sc mappings} photoionization model predictions.  Dividing lines are the same as in Figure~\ref{fig:BPTVO_soft_excess}.  The top row of diagrams is for $p_\mathrm{NT} = 0.1$ and the bottom row is for $p_\mathrm{NT} = 0.4$.  The gray points are observed line ratios for the galaxy NGC~1365 from the S7 galaxy survey \citep{Dopita_2015_S7}, with each point representing the line ratios for the narrowest Gaussian kinematic component in a single spatial pixel.  See Section~\ref{sec:oxaf_predictions} in the text for more information and interpretation of the diagrams.}
	\label{fig:MAPPINGS_grid}
\end{figure*}

A grid of {\sc mappings} photoionization models was used to explore the effect of the {\sc oxaf} parameters on emission-line ratios on the standard optical diagnostic diagrams.  The grid was run over a range of photoionization model parameters and a range of all three {\sc oxaf} parameters.  The results are shown in Figure~\ref{fig:MAPPINGS_grid} for the dusty, plane-parallel {\sc mappings} models configured with metallicity $Z = 1.5$~$Z_\odot$, ionization parameter $U(H) = 10^{-3}$, and a constant pressure of $P/k = 10^7$~K$\,$cm$^{-3}$.  Each model was iteratively re-run three times to ensure that the total pressure (including radiation pressure) was close to the input pressure, with this constraint applied where the model nebula was $50\%$ ionized.  The iterations for each gridpoint culminated in a model which ended when the nebula was $99\%$ neutral.

Also plotted in Figure~\ref{fig:MAPPINGS_grid} are gray points showing the measured spatially-resolved line flux ratios for the galaxy NGC~1365 from the S7 galaxy survey \citep{Dopita_2015_S7}.  Each point represents the line ratios for the narrowest Gaussian kinematic component in the spectrum of a 1"$\times$1" spatial pixel (the seeing FWHM was ${\sim}1.1$").  A S/N cut of 10 was applied to the linear flux ratios.  In the leftmost two columns of diagrams these points form a `mixing sequence' from line-ratios associated with excitation purely due to star formation (at the bottom) to excitation purely by AGN radiation (at the top).

The predicted line ratios for our fiducial {\sc mappings} models approximately coincide with some of the measured `pure-AGN' line ratios, especially in the leftmost column.  Predictions of the weaker [\ion{O}{1}]$\lambda 6300$ line however are inconsistent with the observations.  This line is produced predominantly in the partially ionized zone that arises due to the hard EUV and X-ray photons.  The temperature and size of this region and thus the strength of the [\ion{O}{1}] line are sensitive to the hard ionizing spectrum.  In addition the fractional ionization state of oxygen is strongly dependent upon charge exchange reactions and as a result this line is very difficult to predict accurately.  The [\ion{S}{2}]$\lambda\lambda$~6716,~6731 doublet can also arise from this partially ionized region and is also somewhat inconsistent with the observations, albeit to a lesser degree.  The overpredicted fluxes from these species could to some extent be due to the NLR clouds being mass-bounded as opposed to ionization-bounded (as in the models), i.e.\ more of the high-energy photons escape and the partially-ionized zones are shorter for the observed ensemble of NLR clouds in NGC~1365 compared to the models.


\subsection{Effect of the individual {\sc oxaf} parameters}
In this section we discuss Figure~\ref{fig:MAPPINGS_grid} and interpret the effect of each {\sc oxaf} parameter on the model predictions.

\subsubsection{Energy of the peak of the disk emission $E_\mathrm{peak}$}
The logarithmically-spaced values of $E_\mathrm{peak}$ in Figure~\ref{fig:MAPPINGS_grid} show that increasing the energy of the peak of the disk emission, and thereby increasing the energy of the ionizing photons, increases both the relative ionization state and the temperature of the nebula.  The sensitivity of optical diagnostic line ratios to $E_\mathrm{peak}$ is comparable to the sensitivity to the ionization parameter.

\subsubsection{Proportion of flux in non-thermal component, $p_{NT}$}
We first consider the expected range of values of $p_\mathrm{NT}$.  \citet{Jin_2012_I} study a sample of 51 unobscured Seyfert~1 galaxies and fit the full three-component {\sc optxagnf} model to the observed SEDs.  The proportion of the bolometric luminosity emitted in the power-law component in the best-fit model is less than 0.05 for ${\sim}20\%$ of the sample, less than 0.2 for ${\sim}50\%$ of the sample, and less than 0.4 for ${\sim}80\%$ of the sample.  Although the soft excess is a significant component (allocated more flux than the power-law component for ${\sim}70\%$ of the sample), it is effectively an extension of accretion disk emission to higher energies, so the proportion of flux in the power-law tail is comparable to the $p_\mathrm{NT}$ {\sc oxaf} parameter.  Hence we take $p_\mathrm{NT} \sim 0.1 - 0.4$ as an appropriate range for typical ionizing AGN spectra.

The values $p_\mathrm{NT} = 0.2, 0.7$ were included in the model grid but are not shown in Figure~\ref{fig:MAPPINGS_grid}.  The diagnostic line-ratio predictions for $p_\mathrm{NT} = 0.2$ were intermediate between those of $p_\mathrm{NT} = 0.1$ and $p_\mathrm{NT} = 0.4$ and were similar to those of $p_\mathrm{NT} = 0.1$.  The value $p_\mathrm{NT} = 0.7$ was very high, with only a few extreme objects in the \citet{Jin_2012_I} sample having such power-law-dominated SEDs.  Hence the figure requires only the values $p_\mathrm{NT} = 0.1,$~$0.4$ to indicate the effect of $p_\mathrm{NT}$ and show the approximate magnitude of strong line-ratio variations due to $p_\mathrm{NT}$ variations that may occur between Seyfert galaxies.

Increasing the value of $p_\mathrm{NT}$ increases the length of the trailing partially ionized zone in the photoionization models.  For $E_\mathrm{peak}$~=~32~eV, $\Gamma$~=~1.8, $\log$~U~=~-2.0 and Z~=~1.5~Z$_\odot$, variation of $p_\mathrm{NT}$ from 0.1 to 0.7 increased the length of the modeled cloud by a factor of ${\sim}4$ to 1.5~pc.  Extending the partially-ionized zone increases the [\ion{S}{2}]/H$\alpha$ and [\ion{O}{1}]/H$\alpha$ line ratios, as the figure shows for $p_\mathrm{NT} = 0.1 \rightarrow 0.4$.

\subsubsection{Photon index of the power-law tail $\Gamma$}
For values of $p_\mathrm{NT} \lesssim 0.2$, the predicted emission-line ratios do not show strong sensitivity to $\Gamma$.  Figure~\ref{fig:MAPPINGS_grid} shows how as $\Gamma$ increases, [\ion{S}{2}] and [\ion{O}{1}] line emission is enhanced relative to H$\alpha$.  Increasing $\Gamma$ allocates more of the non-thermal energy to relatively soft X-ray photons which are more easily absorbed by the nebula, causing more emission in the partially ionized zone.  Hence $\Gamma$ has a similar effect to $p_\mathrm{NT}$.  As $p_\mathrm{NT}$ increases, the sensitivity to $\Gamma$ increases due to the greater relative effect of X-ray power-law photons on the nebula ionization structure.

\subsection{Comparison with default {\sc cloudy} AGN spectrum}
\label{sec:oxaf_cloudy}
In this section we compare {\sc oxaf} spectra with the default spectrum produced by the AGN spectral model supplied with the {\sc cloudy} \citep{Ferland_2013_CLOUDY} photoionization code.  We also compare the predicted optical diagnostic emission-line ratios produced by using these ionizing spectra in {\sc MAPPINGS} photoionization models.

The {\sc cloudy} AGN spectral model uses an analytic formula consisting of exponential and power-law factors and terms.  The default {\sc cloudy} AGN model was configured with the following default values: a BBB temperature of $T = 1.5 \times 10^5$~K, an X-ray to UV ratio of $\alpha_{ox} = -1.4$, a low-energy BBB slope of $\alpha_{uv} = -0.5$, and an X-ray power-law index of $\alpha_x = -1$ (corresponding to $\Gamma = 2$).

\begin{figure}
	\centering
	\includegraphics[width=0.48\textwidth]{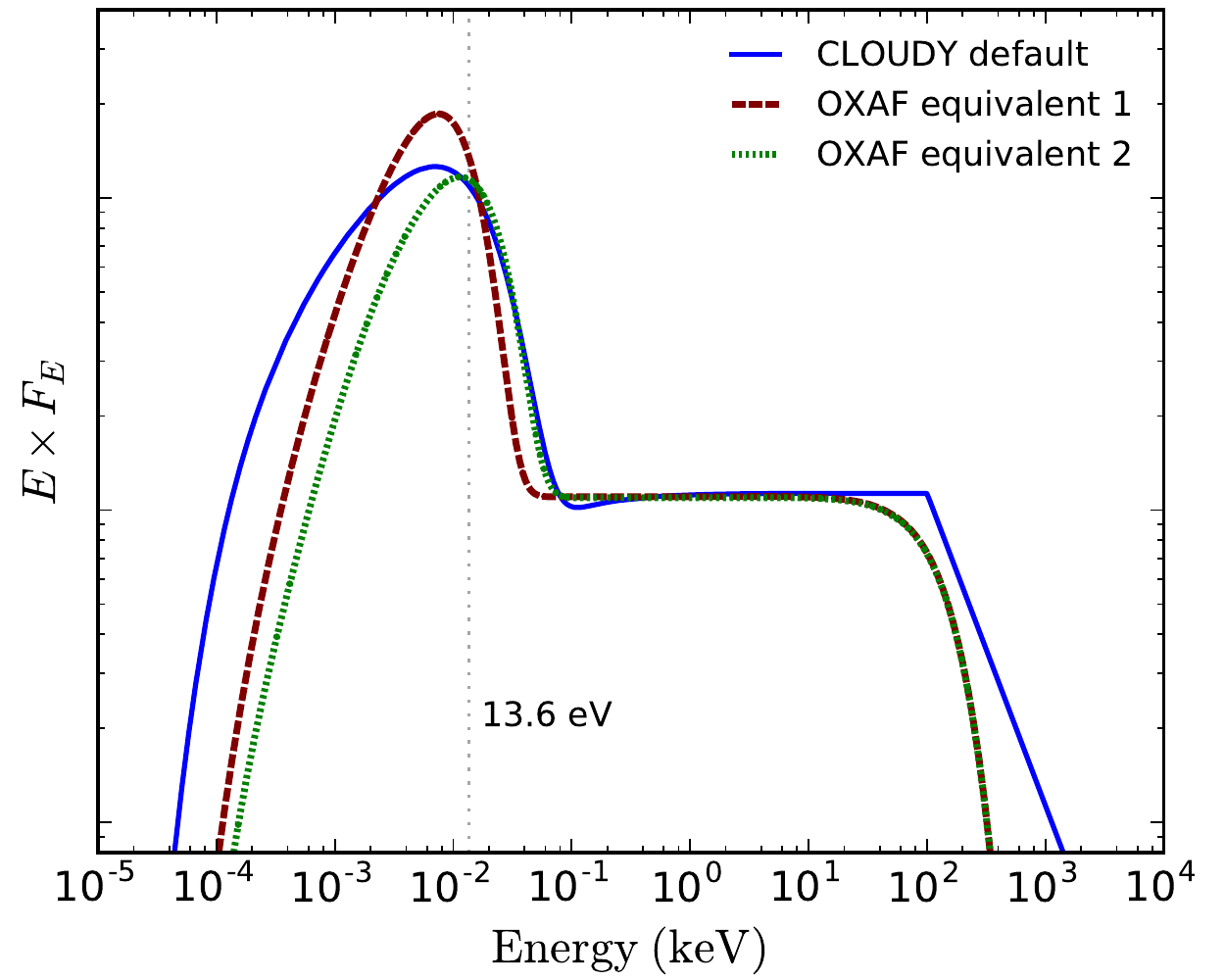}
	\caption{Comparison between a default {\sc cloudy} ionizing AGN spectrum and two {\sc oxaf} spectra chosen to be `equivalent' in differing ways.  The parameters used for the spectra are listed in Section~\ref{sec:oxaf_cloudy}.}
	\label{fig:cloudy_oxaf_spec}
\end{figure}

Figure~\ref{fig:cloudy_oxaf_spec} shows the {\sc cloudy} default SED, along with two `equivalent' {\sc oxaf} spectra.  The first {\sc oxaf} spectrum has $(E_{\mathrm{peak}},$~$\Gamma$,~$p_{\mathrm{NT}}) =$ (6.9~eV,~2.0,~0.46), chosen to match the BBB peak energy and the `height' of the power-law tail, and has the same normalization as the normalized {\sc cloudy} default SED.  The second `equivalent' {\sc oxaf} spectrum has $(E_{\mathrm{peak}},$~$\Gamma$,~$p_{\mathrm{NT}}) =$ (9.9~eV,~2.0,~0.414) to match the shape and position of the high-energy slope of the BBB and match the `height' of the power-law tail.  This second {\sc oxaf} spectrum is shown with a different normalization, with only 73\% of the energy flux of the {\sc cloudy} spectrum, to show how it follows the shape of the {\sc cloudy} spectrum in the important region associated with H-ionizing FUV photons.

Clear differences between the spectral models are apparent in Figure~\ref{fig:cloudy_oxaf_spec}.  The {\sc cloudy} model uses a single blackbody curve combined with a power-law to produce the BBB, resulting in a BBB much wider than the {\sc oxaf} BBB.  The physical model used to calculate the power-law tail in {\sc oxaf} produces a power-law cutoff that is much steeper than the simple piecewise $\nu^{-2}$ cutoff in the {\sc cloudy} model.

\begin{figure*}
	\centering
	\includegraphics[width=1.0\textwidth]{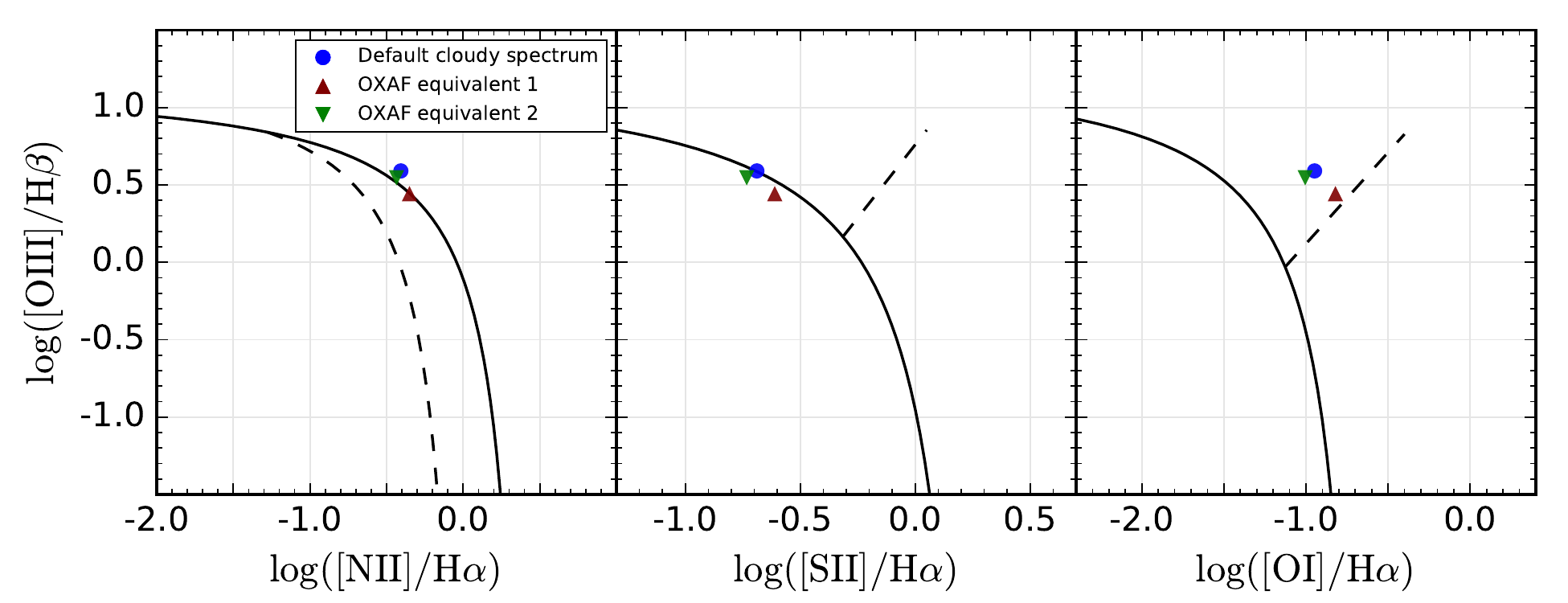}
	\caption{Optical diagnostic diagram showing predicted emission-line flux ratios from {\sc MAPPINGS} photoionization models using the three different ionizing spectra shown in Figure~\ref{fig:cloudy_oxaf_spec}.  The dividing lines are the same as in Figure~\ref{fig:BPTVO_soft_excess}.  The three ionizing spectra are the default {\sc cloudy} AGN spectrum and two {\sc oxaf} spectra that are `equivalent' in differing ways to the {\sc cloudy} spectrum; colors in this figure correspond to the colors of these three spectra in Figure~\ref{fig:cloudy_oxaf_spec}.  The plane-parallel dusty MAPPINGS models were all configured with a metallicity of $3 \, Z_\odot$, an ionization parameter of $\log U = -2.0$, and a constant pressure of $P/k = 10^6$. \vspace{0.8cm}}
	\label{fig:cloudy_oxaf_BPT}
\end{figure*}

Results of {\sc MAPPINGS} photoionization model runs using the three spectra in Figure~\ref{fig:cloudy_oxaf_spec} are presented in Figure~\ref{fig:cloudy_oxaf_BPT}.  The plane-parallel dusty MAPPINGS models were all configured with a metallicity of $3 \, Z_\odot$, an ionization parameter of $\log U = -2.0$, and a constant pressure of $P/k = 10^6$.  All three models produced similar optical diagnostic line ratios.  The second {\sc oxaf} model resulted in line ratios closer to those due to the {\sc cloudy} default spectrum, which confirms that the most important part of the spectrum for ionization modeling is the region immediately on the high-energy side of the H ionization threshold, where these spectra were very similar by design.  The first {\sc oxaf} model results in relatively lower [\ion{O}{3}]$/$H$\beta$, which is presumably a consequence of the lower proportion of hydrogen-ionizing EUV photons.

The results show that {\sc oxaf} spectra may produce predicted diagnostic line ratios that are very similar to those resulting from analytic ionizing spectra not based on physical models, provided the spectral shapes are sufficiently well-matched in the most important part of the spectrum, the hydrogen-ionizing EUV.

\section{Conclusions}
\label{sec:conclusion}
We present a model of AGN continuum emission, {\sc oxaf}, designed for use in photoionization modeling and featuring spectral shapes based on physical models.  The {\sc oxaf} model removes degeneracies between AGN parameters in terms of their effect on the ionizing spectral shape; in particular, in the thin disk accretion model the black hole mass and AGN luminosity are entirely degenerate with respect to their impact on the spectral shape.  We base {\sc oxaf} on the {\sc optxagnf} model \citep{Done_2012_AGN_SED}, and explain and remove some additional spectral-shape degeneracies between AGN parameters when we reparametrize {\sc optxagnf}.  

We show that the soft X-ray excess is not important to photoionization modeling of the standard optical strong diagnostic lines, so we do not include it in the {\sc oxaf} model.  However the soft excess must be important in modeling forbidden high-ionization lines.

The model {\sc oxaf} contains only three parameters - the energy of the peak of the accretion disk emission $E_{\mathrm{peak}}$, the photon power-law index for the non-thermal Comptonized component $\Gamma$, and the proportion of the total flux which goes to the non-thermal component $p_\mathrm{NT}$.  These parameters intuitively describe the physical shape of the produced spectrum.

We show that predicted ratios of strong lines on standard optical diagnostic diagrams are sensitive to all three {\sc oxaf} parameters.  The parameter $E_{\mathrm{peak}}$ directly affects the degree of ionization of the {\sc mappings} model nebulae.  The parameters $\Gamma$ and $p_\mathrm{NT}$ are similar in their effects in that they change the length of the partially-ionized zone.  Predicted line-ratios are more sensitive to $\Gamma$ as $p_\mathrm{NT}$ is increased.  Measured strong emission-line ratios for the Seyfert galaxy NGC~1365 are approximately consistent with the predictions of some fiducial photoionization models using input {\sc oxaf} ionizing spectra.

Users of {\sc oxaf} may further explore the effects of changing the shape of the ionizing spectrum on predicted emission-line spectra.

\acknowledgments
B.G. gratefully acknowledges the support of the Australian Research Council as the recipient of a Future Fellowship (FT140101202).  L.K. and M.D. acknowledge support from ARC discovery project \#DP160103631.  C.J. acknowledges the support by the Bundesministerium f\"{u}r Wirtschaft und Technologie/Deutsches Zentrum f\"{u}r Luft- und Raumfahrt (BMWI/DLR, FKZ 50 OR 1604) and the Max Planck Society.  We are grateful to both Chris Done, who received an early draft of this paper, and the anonymous referee for helpful comments.



\begin{figure}
  \centering
      \includegraphics[width=1\textwidth]{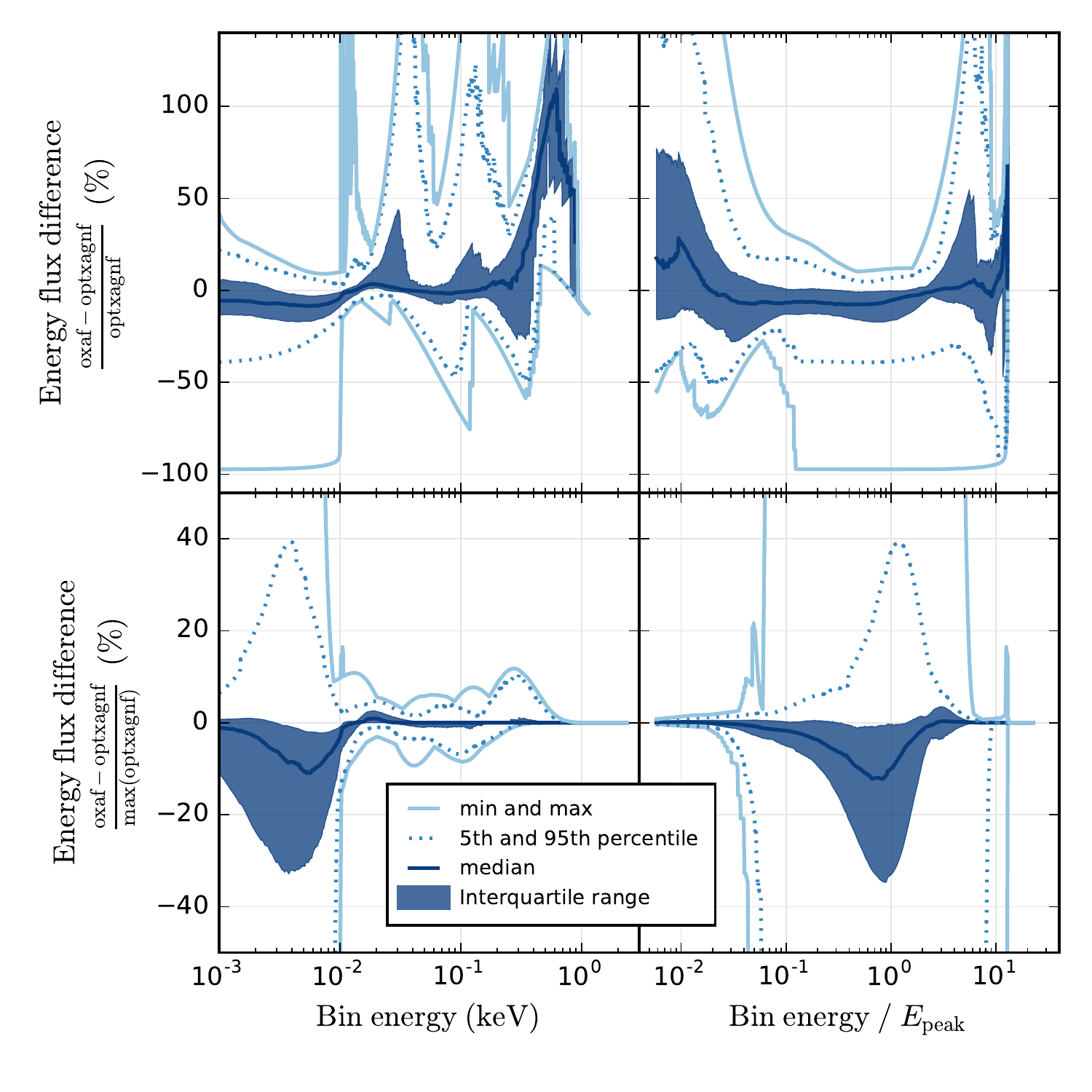}
  \caption{Comparison of the predicted Big Blue Bump (BBB) disk emission between the {\sc oxaf} and {\sc optxagnf} models.  For each spectrum the predicted total {\sc oxaf} flux in each bin minus the predicted total {\sc optxagnf} flux in each bin is calculated.  In the subplots in the top row, this difference is reported as a percentage of the {\sc optxagnf} flux in each bin; in the subplots in the bottom row, the difference is reported as a percentage of the maximum predicted {\sc optxagnf} bin flux in the energy range $0.01 - 20$~keV.  The statistics in each spectral bin were taken over {\sc oxaf}-{\sc optxagnf} comparisons at ${\sim}1900$ points in the parameter space, and are plotted as a function of bin energy (left column), and bin energy normalized by the energy of the peak of the disk emission $E_{\mathrm{peak}}$ (right column).  For subplots in the top row, a bin in a spectrum was only included in the statistics if the {\sc optxagnf} bin flux was more than 0.1\% of the maximum {\sc optxagnf} flux for a bin in the range $0.01 - 20$~keV.  The accuracy of {\sc oxaf} in reproducing the {\sc optxagnf} spectra is more than adequate over the relevant energy range of $0.01 - {\sim}20$~keV. }
  \label{fig:oxaf_validation_BBB}
\end{figure}
\begin{figure}
  \centering
      \includegraphics[width=1\textwidth]{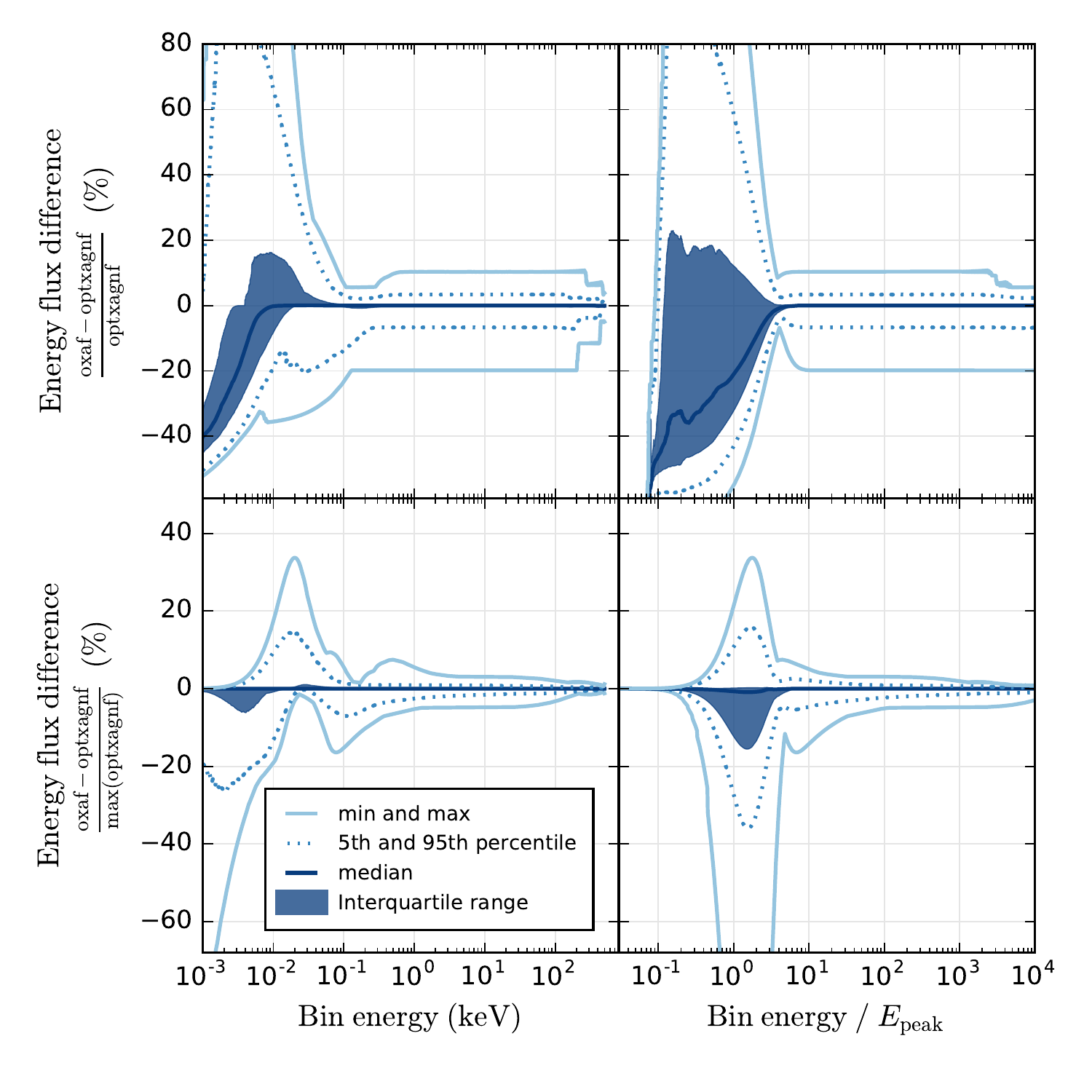}
  \caption{Comparison of the predicted hard non-thermal emission between the {\sc oxaf} and {\sc optxagnf} models.  For each spectrum we calculate the predicted total {\sc oxaf} flux in each bin minus the predicted total {\sc optxagnf} flux in each bin.  In the subplots in the top row, this difference is reported as a percentage of the {\sc optxagnf} flux in each bin; in the subplots in the bottom row, the difference is reported as a percentage of the maximum predicted {\sc optxagnf} bin flux in the energy range $0.01 - 20$~keV.  The statistics in each spectral bin were taken over {\sc oxaf}-{\sc optxagnf} comparisons at ${\sim}1900$ points in the parameter space, and are plotted as a function of bin energy (left column), and bin energy normalized by the energy of the peak of the disk emission $E_{\mathrm{peak}}$ (right column).  For subplots in the top row, a bin in a spectrum was only included in the statistics if the {\sc optxagnf} bin flux was more than 0.1\% of the maximum {\sc optxagnf} flux for a bin in the range $0.01 - 20$~keV.  The accuracy of {\sc oxaf} in reproducing the {\sc optxagnf} spectra is more than adequate over the relevant energy range of $0.01 - {\sim}20$~keV.}
  \label{fig:oxaf_validation_NT}
\end{figure}

\appendix
Plots comparing {\sc oxaf} with {\sc optxagnf} are presented in Figures~\ref{fig:oxaf_validation_BBB} and \ref{fig:oxaf_validation_NT}.

\begin{thebibliography}{}
\bibitem[Allen et al.(1998)]{Allen_1998_NLRs_UV} Allen, M.~G., Dopita, M.~A., \& Tsvetanov, Z.~I.\ 1998, \apj, 493, 571 
\bibitem[Antonucci (1993)]{Antonucci_1993_AGN_unified} Antonucci, R.\ 1993, \araa, 31, 473
\bibitem[Baldwin, Phillips and Terlevich (1981)]{BPT_1981} Baldwin, J.~A., Phillips, M.~M., \& Terlevich, R.\ 1981, \pasp, 93, 5
\bibitem[Bianchi et al.(2006)]{Bianchi_2006_xray_excess} Bianchi, S., Guainazzi, M., \& Chiaberge, M.  2006, \aap, 448, 499
\bibitem[Bland-Hawthorn et al.(2013)]{Bland-Hawthorn_2013_MW_centre} Bland-Hawthorn, J., Maloney, P.~R., Sutherland, R.~S., \& Madsen, G.~J.\ 2013, \apj , 778, 58 
\bibitem[Collins et al.(2009)]{Collins_2009} Collins, N.~R., Kraemer, S.~B., Crenshaw, D.~M., Bruhweiler, F.~C., \& Mel{\'e}ndez, M.\ 2009, \apj, 694, 765 
\bibitem[Contini \& Viegas(2001)]{Contini_Viegas_2001} Contini, M., \& Viegas, S.~M.\ 2001, \apjs, 132, 211
\bibitem[Czerny et al.(2016)]{Czerny_2016} Czerny, B., You, B., Kurcz, A., et al.\ 2016, arXiv:1601.02498
\bibitem[Davies et al.(2016)]{Davies_2016_S7_NLR_rad_P} Davies, R.~L., Dopita, M.~A., Kewley, L., et al.\ 2016, \apj, 824, 50
\bibitem[Davis et al.(2006)]{Davis_Done_Blaes_2006} Davis, S.~W., Done, C., \& Blaes, O.~M.\ 2006, \apj, 647, 525
\bibitem[Done et al.(2012)]{Done_2012_AGN_SED} Done, C., Davis, S. W., Jin, C., Blaes, O., \& Ward, M.  2012, \mnras , 420, 1848
\bibitem[De Marco et al.(2013)]{De_Marco_2013_AGN_X-ray_lags} De Marco, B., Ponti, G., Cappi, M., et al.\ 2013, \mnras , 431, 2441
\bibitem[Dopita et al.(2002)]{Dopita_2002_Dusty_NLRs} Dopita, M.~A., Groves, B.~A., Sutherland, R.~S., Binette, L., \& Cecil, G.\ 2002, \apj, 572, 753
\bibitem[Dopita et al.(2014)]{Dopita_2014_S7_I} Dopita, M.~A., Scharw{\"a}chter, J., Shastri, P., et al.\ 2014, \aap , 566, A41
\bibitem[Dopita et al.(2015)]{Dopita_2015_S7} Dopita, M.~A., Shastri, P., Davies, R.~L., et al.\ 2015, \apjs, 217, 12 
\bibitem[Evans et al.(1999)]{Evans_1999} Evans, I., Koratkar, A., Allen, M., Dopita, M., \& Tsvetanov, Z.\ 1999, \apj, 521, 531
\bibitem[Fabian(2016)]{Fabian_2016_BH_accretion} Fabian, A.~C.\ 2016, Astronomische Nachrichten, 337, 375
\bibitem[Ferland et al.(2013)]{Ferland_2013_CLOUDY} Ferland, G.~J., Porter, R.~L., van Hoof, P.~A.~M., et al.\ 2013, RMxAA, 49, 137
\bibitem[Gierli{\'n}ski \& Done(2004)]{Gierliski_Done_2004} Gierli{\'n}ski, M., \& Done, C.\ 2004, \mnras, 349, L7
\bibitem[Groves et al.(2004)]{Groves_2004_NLR_I} Groves, B.~A., Dopita, M.~A., \& Sutherland, R.~S.\ 2004, \apjs, 153, 9
\bibitem[Groves et al.(2006)]{Groves_2006_low_Z_AGN} Groves, B.~A., Heckman, T.~M., \& Kauffmann, G.\ 2006, \mnras, 371, 1559
\bibitem[Jin et al.(2012a)]{Jin_2012_I} Jin, C., Ward, M., Done, C., \& Gelbord, J.\ 2012, \mnras, 420, 1825
\bibitem[Jin et al.(2012b)]{Jin_2012_II} Jin, C., Ward, M., \& Done, C.\ 2012, \mnras, 422, 3268 
\bibitem[Jin et al.(2012c)]{Jin_2012_III} Jin, C., Ward, M., \& Done, C.\ 2012, \mnras, 425, 907 
\bibitem[Kauffmann et al.(2003)]{Kauffmann_2003_AGN} Kauffmann, G., Heckman, T.~M., Tremonti, C., et al.\ 2003, \mnras, 346, 1055
\bibitem[Kewley et al.(2001)]{Kewley_2001_starburst} Kewley, L.~J., Dopita, M.~A., Sutherland, R.~S., Heisler, C.~A., \& Trevena, J.\ 2001, \apj, 556, 121
\bibitem[Kewley et al.(2006)]{Kewley_2006_AGN_hosts} Kewley, L.~J., Groves, B., Kauffmann, G., \& Heckman, T.\ 2006, \mnras, 372, 961
\bibitem[Laor \& Netzer(1989)]{Laor_Netzer_1989_accretion_disks} Laor, A., \& Netzer, H.\ 1989, \mnras, 238, 897 
\bibitem[Lightman \& Zdziarski(1987)]{Lightman_Zdziarski_1987} Lightman, A.~P., \& Zdziarski, A.~A.\ 1987, \apj, 319, 643
\bibitem[Molina et al.(2013)]{2013_IBIS_Xray_AGN} Molina, M., Bassani, L., Malizia, A., et al.\ 2013, \mnras, 433, 1687 
\bibitem[Murayama \& Taniguchi(1998)]{Murayama_Taniguchi_1998} Murayama, T., \& Taniguchi, Y.\ 1998, \apjl, 503, L115
\bibitem[Novikov \& Thorne(1973)]{Novikov_Thorne_1973} Novikov, I.~D., \& Thorne, K.~S.  1973, Black Holes (Les Astres Occlus), 343 
\bibitem[Page \& Thorne(1974)]{Page_Thorne_1974} Page, D.~N., \& Thorne, K.~S.  1974, \apj , 191, 499 
\bibitem[Pal et al.(2016)]{Pal_2016_X-ray_AGN} Pal, M., Dewangan, G.~C., Misra, R., \& Pawar, P.~K.  2016, \mnras , 457, 875
\bibitem[Reynolds(2014)]{Reynolds_2014_BH_spin} Reynolds, C.~S.\ 2014, \ssr, 183, 277
\bibitem[Shakura \& Sunyaev(1973)]{Shakura_Sunyaev_1973} Shakura, N.~I., \& Sunyaev, R.~A.\ 1973, \aap, 24, 337
\bibitem[Veilleux \& Osterbrock (1987)]{1987VO} Veilleux, S., \& Osterbrock, D.~E.\ 1987, \apjs, 63, 295
\bibitem[Viegas-Aldrovandi \& Contini(1989)]{Viegas-Aldrovandi_Contini_1989} Viegas-Aldrovandi, S.~M., \& Contini, M.\ 1989, \aap, 215, 253
\bibitem[Vika et al.(2009)]{2009_Vika_SMBH_M_func} Vika, M., Driver, S.~P., Graham, A.~W., \& Liske, J.\ 2009, \mnras, 400, 1451
\bibitem[Walton et al.(2013)]{Walton_2013_reflection} Walton, D.~J., Nardini, E., Fabian, A.~C., Gallo, L.~C., \& Reis, R.~C.\ 2013, \mnras, 428, 2901
\bibitem[You et al.(2012)]{You_2012_GR_BH_spectrum} You, B., Cao, X., \& Yuan, Y.-F.\ 2012, \apj, 761, 109
\bibitem[Zdziarski et al.(1996)]{Zdziarski_1996} Zdziarski, A.~A., Johnson, W.~N., \& Magdziarz, P.\ 1996, \mnras, 283, 193 
\bibitem[{\.Z}ycki et al.(1999)]{Zycki_1999} {\.Z}ycki, P.~T., Done, C., \& Smith, D.~A.\ 1999, \mnras, 309, 561
\end{thebibliography}
\end{document}